\documentclass[
 reprint,
 amsmath,amssymb,
 nofootinbib,aps,pra,floatfix
]{revtex4-2}

\usepackage{acronym} 
\usepackage{appendix}
\usepackage{physics}
\usepackage{siunitx}
\usepackage{bm}
\usepackage{braket}
\usepackage[colorlinks=true, citecolor=blue, linkcolor=blue, urlcolor=blue]{hyperref}
\usepackage{dcolumn}
\usepackage{enumitem}
\usepackage{graphicx}
\usepackage{stmaryrd}
\usepackage{soul}
\usepackage{upgreek} 
\graphicspath{ {FIGURES/} }

\newcommand{\kett}[2][+]{\ket{#1\frac{#2}{2}}_{g}}
\newcommand{\kette}[2][+]{\ket{#1\frac{#2}{2}}_{e}}
\newcommand{\trans}[3][+]{\kett[#1]{#2} \rightarrow \kette[#1]{#3}}
\newcommand{\isotope}[2][167]{$^{#1}$#2$^{3+}$:Y$_2$SiO$_5$}

\newcommand{\YSO}{Y$_2$SiO$_5$}
\newcommand{\us}{$\upmu$s}
\newcommand{\um}{$\upmu$m}

\newacro{rase} [RASE] {Rephased Amplified Spontaneous Emission}
\newacro{ase} [ASE] {Amplified Spontaneous Emission}
\newacro{fft} [FFT] {fast Fourier transform}

\begin{document}

\title{High performance spin-state memory for telecom quantum information applications}
\author{James Stuart}
\affiliation{Research School of Physics, Australian National University, Canberra, ACT, Australia}
\author{Kieran Smith}
\affiliation{Research School of Physics, Australian National University, Canberra, ACT, Australia}
\author{Morgan Hedges}
\affiliation{Research School of Physics, Australian National University, Canberra, ACT, Australia}
\author{Rose Ahlefeldt}
\affiliation{Research School of Physics, Australian National University, Canberra, ACT, Australia}
\author{Matthew Sellars}
\affiliation{Research School of Physics, Australian National University, Canberra, ACT, Australia}

\begin{abstract}
    Optical quantum memories based on Er, with its telecom-compatible optical transition, could benefit quantum networking and linear optics quantum computing applications, if the requirements for high memory efficiency, fidelity, and storage capacity can be met simultaneously. Here, we demonstrate the first high efficiency  memory in Er, with spin-state storage for 25~\us\ and efficiencies up to 81\%, using the rephased amplified spontaneous emission protocol. We show high-fidelity entanglement of an output state and the state stored in memory, and a demonstration of multiplexing, storing 70 modes in the temporal domain. We use the results to discuss how, with modifications to the pulse sequences and experimental geometry only, the performance can be improved to meet the requirements for practical quantum applications. 
\end{abstract}

\maketitle

\section{Introduction}
    Quantum memories are a critical component for optical quantum information processing, allowing the synchronisation of non-deterministic events. In this role quantum memories have been proposed for applications such as quantum repeaters \cite{Sangouard11, Tittel10}, circuit-based linear optics quantum computing, and in cluster state generation \cite{Choi19, Manukhova17} for measurement-based quantum computing \cite{Raussendorf01, Nielsen04}. Beyond this synchronisation role, recent work has suggested that integrating memories into quantum computers may reduce physical qubit requirements for fault tolerant operation 100-fold \cite{mundada26}. Further, memories can be used to perform other key processing operations such as mode switching, mode mixing, and entanglement generation \cite{moiseev16, hosseini12, leung_observation_2022, Yokoyama13, muralidharan16}. Protocols already exist for combining some of these tasks in a single element, such as combining entanglement generation and quantum storage for quantum repeaters\cite{duan_long-distance_2001, Ledingham10}, minimizing the losses associated with passing information into and out of multiple discrete components. Quantum networks are likely to benefit from even more complex multi-tasking components, for tasks like routing. Such devices are essentially small linear optics quantum computers (LOQC) and there is, indeed, the potential for quantum memories to reduce the complexity and losses currently encountered in scaling current photonic quantum processors, whilst greatly increasing the time quantum information can be held within a calculation. 
    
     The widespread uptake of quantum memories for quantum information processing in computers and networks is awaiting the development of a quantum memory technology with high performance along with wavelength compatibility with existing optical technology and, ideally, the ability to multi-task as described above. Key performance metrics are the memory efficiency, fidelity, data storage capacity, and storage time. For instance, global scale quantum repeater networks would require memory components with efficiencies above 90\% \cite{Tittel10,muralidharan16}, near unity fidelity, 100~ms storage times \cite{razavi_quantum_2009} and compatibility with fibre telecommunications bands. No fixed thresholds have been set for data capacity; however, MHz data rates would require 100,000 modes, for a global network. More general linear optics quantum processing applications require similar performance, albeit with lower storage times. These metrics have proven challenging to optimise individually, and extremely difficult to optimise simultaneously. For instance, a small handful of quantum memories have shown efficiencies above 70\% \cite{hedges10,Schraft16,Hosseini11, Hsiao18,cao20,wang19,guo25}, reaching 94.6\% (0.24 dB loss) for Rb gas ensembles \cite{guo25}. This latter impressive result is already comparable with the storage efficiencies of the delay lines currently used for quantum storage in commercial LOQC systems (e.g. typical $>0.1$ dB chip-to-fiber loss \cite{alexander_manufacturable_2025}), but these systems are not readily solid-state or telecom compatible, and their material properties limit storage time and mode capacity.
     $^{167}$Er-based crystals represent an attractive class of candidates where optical performance may be possible across all performance metrics for optical quantum processing applications. $^{167}$Er has an optical transition in the telecom {C}-band, where the lowest-loss photonic components are available. Long coherence times can be achieved on its optical (4~ms \cite{Bottger73Hz}) and hyperfine (1.3~s \cite{Rancic16}) transitions by choosing low decoherence materials such as \YSO\ and applying some combination of high magnetic fields and low temperatures. $^{167}$Er:\YSO, specifically, has small optical inhomogeneous linewidths (150~MHz) compared to its $\mathcal{O}$(GHz) hyperfine splittings (ranging between 790 - 1,000 MHz), which, coupled with its long hyperfine lifetime (180~s \cite{Stuart21}) in high fields,  allows spectral preparation of the ensemble into a pure state \cite{Rancic16, Stuart21}. The complete control over the atomic state offered by this feature allows high bandwidth operation (for high storage capacity across spectro-temporal degrees of freedom) while retaining access to long hyperfine storage times, two characteristics otherwise traded off in materials without pure state preparation.

A small number of quantum memory protocols have been demonstrated to date in \isotope{Er}:  a non-classical two-level atomic frequency comb (AFC) with an efficiency of 22\% \cite{Stuart21} and two classical two-level revival of silenced echo (ROSE \cite{damon11}) demonstrations with efficiencies up to 44\% \cite{Minnegaliev22, Minnegaliev23}. These results highlight the potential of $^{167}$Er, as observed efficiencies were constrained by technical challenges rather than intrinsic material properties. However, these existing demonstrations did not employ a memory protocol capable of simultaneous high efficiency, fidelity, storage time, and storage capacity, as required for repeater and general LOQC applications. Here, we implement a memory protocol with the potential to meet these requirements, 4-level rephased amplified spontaneous emission (RASE) \cite{Ledingham10, Beavan12, kate16}, in \isotope{Er}, demonstrating simultaneous storage of tens of modes with efficiencies up to 81\% and low noise.
    
\section{Experiment}
    \subsection{Material and spectral preparation}
    All experiments in this work used the 1538~nm transition between the lowest doublets of the $^4$I$_{15/2}$ ground and $^4$I$_{13/2}$ excited state in site 2 of a 0.005\% \isotope{Er} crystal, grown by Scientific Materials, with an isotopic purity of 92\%. The crystal had dimensions of $3 \times 4 \times 5$~mm cut along the optical extinction axes ($D_1\times D_2 \times b$) \cite{li_spectroscopic_1992}. To suppress electron spin flips, which lead to excessive decoherence and short hyperfine lifetimes in Er crystals, we used the regime identified in earlier work \cite{Rancic16}: the crystal was cooled to 1.5 K in a pumped liquid helium cryostat and a magnetic field of 6 T was applied. This field was aligned $\sim 1^\circ$ off the crystal's $D_1$ axis to spectrally separate the magnetically inequivalent subgroups of Er ions \cite{Sun08}. All experiments were then performed on a single subgroup. Optical access to the crystal was only available through a single window at the bottom of the cryostat. Light entered via this window, propagating $\sim 1^\circ$ off the $D_1$ axis, and was reflected through the crystal by a mirror. All measurements were made in this double pass configuration, with the small angle separating the two beams chosen to avoid optical standing waves across the crystal.

    In the above regime, the optical transition chosen has an optical coherence time of 1.35 ms \cite{milosThesis} and a ground state hyperfine coherence time of 1.3 seconds \cite{Rancic16}(nuclear spin 7/2). Further, the 180~s hyperfine lifetime means the population can be readily spectrally prepared into very high optical depth features on minimal absorption backgrounds via optical pumping. The preparation used here involves first optically pumping the entire ensemble into a single hyperfine state ($\ket{- \frac{7}{2}}$) \cite{Rancic16,Stuart21}. Then, an $\mathcal{O}$(MHz) sub-ensemble is pumped into the opposite hyperfine state ($\ket{+ \frac{7}{2}}$) \cite{Stuart21}, achieving a large spectral and angular momentum separation between this memory feature and the bath of unused Er ions that limits relaxation pathways for the memory ions. 

    Using a similar magnetic field, temperature and spectral preparation process, a two-level atomic frequency comb (AFC) \cite{Stuart21} was previously demonstrated with an efficiency of 22\%,  almost two orders of magnitude higher than earlier, low-field memory demonstrations in Er \cite{Saglamyurek15, Lauritzen2009}. The 22\% efficiency was limited by background absorption at the memory frequency. Here, this issue has been addressed by improving the laser frequency stability to allow better control of the population during spectral preparation. 
 
    \subsection{Memory protocol}    
    4-level RASE \cite{Beavan12, kate16, Duda23} is a spin-state quantum memory protocol that creates two entangled, time-separated optical states \cite{Beavan12, kate16}, where the second optical state is stored and recalled from the ensemble. The 4-level RASE protocol is, thus, a source of entangled light states, allowing it to replace both the quantum light source and the quantum memory components of a quantum network or other LOQC component.
  
\begin{figure}
            \centering
            \includegraphics[width=\linewidth]{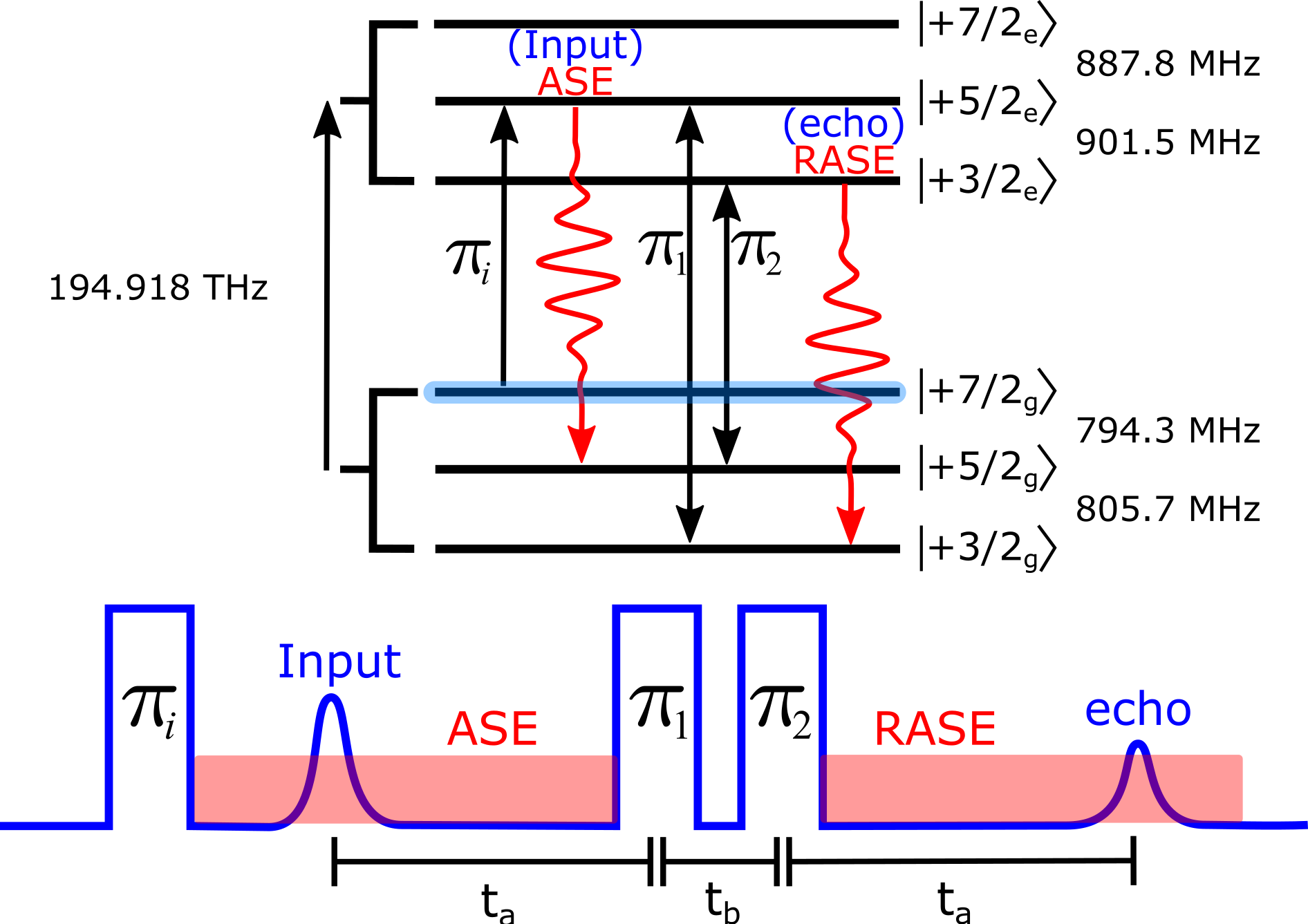}
            \caption{4-level RASE protocol. \textbf{Top:} level diagram indicating the optical transitions used for RASE. The ensemble was spin polarized into the $\kett[-]{7}$ and a narrow, MHz wide, feature was prepared into the $\kett[+]{7}$ hyperfine level (highlighted in blue). 
            \textbf{Bottom:} RASE pulse sequence with an added input pulse to provide an echo (I4LE). The separation between $\pi_1$ and $\pi_2$ defines the memory storage time $t_b$. The gain on the ASE transition can be controlled by attenuating $\pi_i$.}
            \label{fig:level_diagram}
        \end{figure}

    The 4-level RASE protocol is depicted in Figure \ref{fig:level_diagram}. The first output state is amplified spontaneous emission, ASE, generated from an ensemble inverted by $\pi_i$. The ASE photonic state is entangled with the atomic state on the optical levels of the Er ions. This atomic state is then stored on the hyperfine-levels using a $\pi$-pulse ($\pi_1$), and then recalled after a programmable time delay ($t_b$) by a second $\pi$-pulse ($\pi_2$). The resulting state emitted from the ensemble is rephased amplified spontaneous emission, RASE, time-symmetric with the original ASE state. The 4-level version of RASE has the advantage over the earlier, 2-level RASE protocol \cite{Beavan12,Ledingham12} of reduced noise, since all control pulses are spectrally resolved from the ASE and RASE fields, and hyperfine storage \cite{kate16}. Here, the storage time is limited to $\mathcal{O}$(10 \us), due to inhomogeneity on the hyperfine transition, but can be extended to the 1.3 s hyperfine coherence time by using hyperfine rephasing pulses.
    
    Figure \ref{fig:level_diagram} also depicts the inverted-four-level echo (I4LE) protocol \cite{Duda23}, which can be implemented at the same time as RASE by probing the ASE transition with a weak, coherent pulse, a method used here to characterize the memory. In this case, the ensemble emits both RASE and an echo of the probe \cite{Duda23}. RASE itself is not a read-write quantum memory because it does not have an input, and due to the gain arising from the inversion, I4LE is not suitable for quantum storage. However, the 4-level RASE protocol can be transformed into the noiseless photon echo (NLPE) protocol \cite{Ma21}, a read-write memory, with only minor changes to the pulse sequence in Figure \ref{fig:level_diagram} (see Sec. \ref{sec:discussion}). Thus, the results here inform on the performance of read-write quantum memories in \isotope{Er} as well.

        RASE was first demonstrated in Pr:\YSO~using the two-level protocol~\cite{Ledingham10}, followed by demonstrations of the 4-level RASE and I4LE protocols in the same material \cite{Beavan12, kate16, Duda23}. Non-classical correlations were observed \cite{kate16}, with optimised recall efficiencies of 14\% \cite{Duda23}. In both demonstrations, the low recall efficiency was identified as due to the distortion of the rephasing $\pi$-pulses at higher optical depths, so the peak efficiency was reached at a low optical depth of $\alpha L \approx 1$ \cite{Duda23}. Pulse distortion arises when bright pulses are applied on transitions with high oscillator strengths, unavoidable in Pr:\YSO\ since the only available transitions for the $\pi$-pulses have similar oscillator strengths to the ASE and RASE transitions. This is a fundamental limitation on the efficiency of echo-based protocols in Pr:\YSO~in a free-space configuration.

      There is some prospect of improving the low rephasing efficiency of Pr:\YSO\ using cavity enhancement \cite{williamson14}, but the visible operation frequency (606 nm), small hyperfine structure ($\mathcal{O}$(10~MHz)) and thus bandwidth, and large inhomogeneous broadening ($\mathcal{O}$(10 GHz)) are also not well suited for quantum information applications. These reasons motivated the choice here of $^{167}$Er:\YSO. It has more favorable oscillator strength ratios, explained further in Section \ref{sec:results:i4le}, and much narrower inhomogeneous linewidths than Pr:\YSO. The large, resolved hyperfine structure in the high field regime has several benefits for memories: higher bandwidth operation and better spectral preparation, as already discussed, along with much better spectral isolation of quantum memory features, reducing loss via background absorption. We previously reported an optimised spectral preparation process for quantum memories in $^{167}$Er:\YSO \cite{Stuart21} that makes best use of these features; here, we concentrate on the implementation of the RASE protocol and what information it provides on the quantum memory performance possible in future devices.
        
        

    \subsection{Experimental setup}
        \begin{figure}
            \centering
            \includegraphics[width=\linewidth]{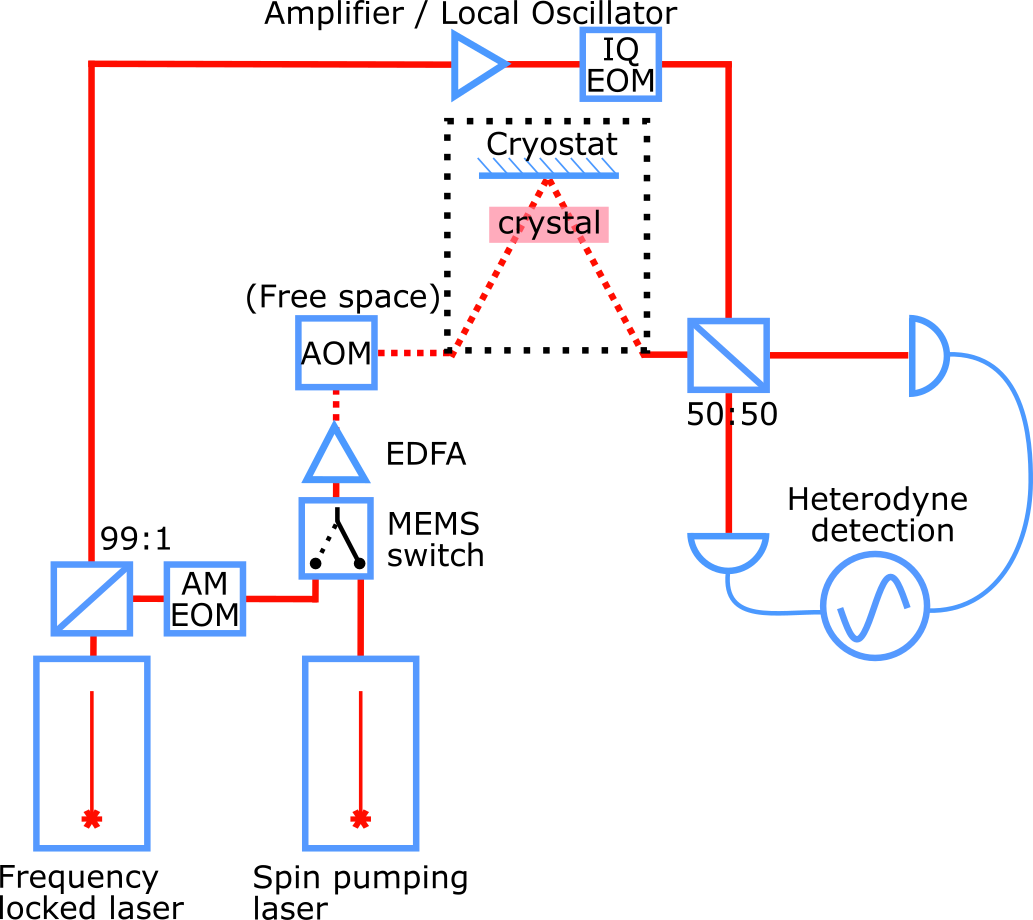}
            \caption{Experiment setup. Two lasers were used: a frequency locked laser for the RASE sequence, and a spin-pumping laser for spectral preparation into a single hyperfine state \cite{Stuart21}. The free space acousto-optic modulator (AOM) was used to gate ASE produced by the Er-doped fiber amplifier (EDFA) during the RASE experiments. Otherwise, broadband ASE produced by the EDFA drives the ASE transitions and acts as background noise in the RASE window. The local oscillator was an amplified pick-off of the frequency-locked laser passed through an in-phase and quadrature electro-optic modulator (IQ EOM) that was operating in carrier-suppressed single-sideband mode. Operating this way meant only one sideband from the amplitude-modulating EOM (AM EOM) was detected. }
            \label{fig:setup}
        \end{figure}
        The experimental setup, Figure \ref{fig:setup}, was similar to Ref. \cite{Stuart21}. For each experiment performed, the ensemble was spin polarised into the $\kett[-]{7}$ hyperfine level and then a 1 MHz wide sub-ensemble was prepared into the $\kett{7}$ hyperfine level.
    
        Several practical considerations influenced the choice of pulse intensities and timing of the RASE pulse sequence, Figure \ref{fig:level_diagram}. First, the gain on the ASE transition, which could be controlled by changing the amplitude of $\pi_i$ at fixed duration, reached a maximum of $\approx 36$ dB when $\pi_i$ was a $\pi$-pulse. At this high gain, the I4LE input pulses had to be kept weak to avoid over-driving the ensemble; to meet this condition, we used $\ll 0.1$\% of a $\pi$-pulse. Second, the time delay between the input and rephasing pulses, $t_a$, was chosen to be long enough ($>10$ \us) that an ASE signal could be detected between the  I4LE input pulse and the rephasing $\pi$-pulses, except for the results in Section \ref{sec:results:i4le}. 
        \begin{figure}
            \centering
            \includegraphics[width=\linewidth]{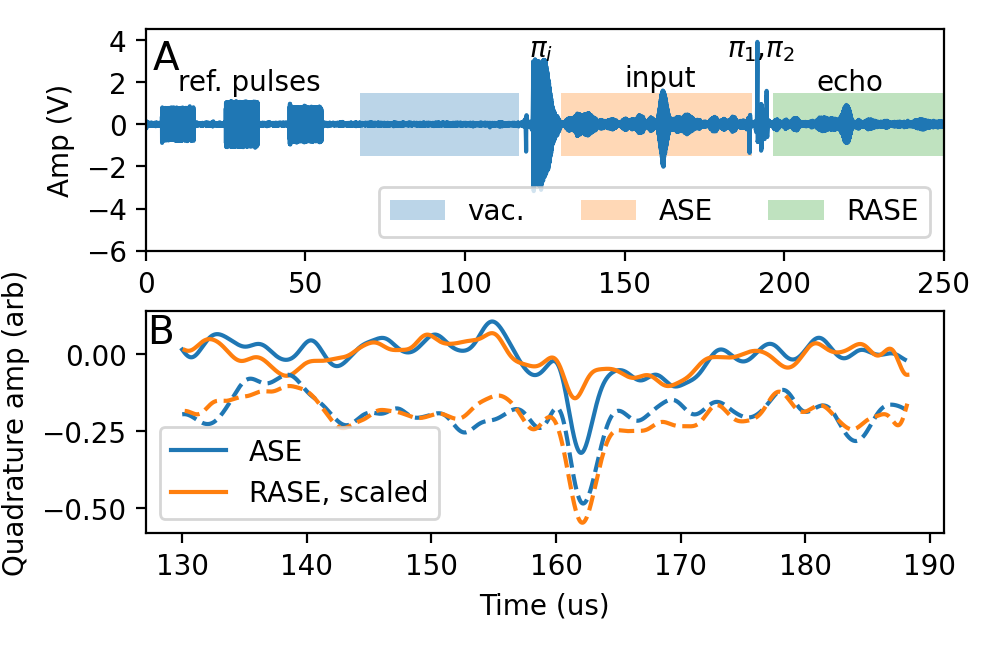}
            \caption{\textbf{A}. High gain I4LE time trace illustrating the experimental pulse sequence, with the windows used for ASE, RASE, and the vacuum reference highlighted. Three phase-correcting reference pulses were used to account for phase and timing jitter between experimental shots.
            \textbf{B}. Comparison of ASE and RASE fields for the high gain data. The fields have been digitally beaten to homodyne and the RASE field has been transformed to match the ASE field (Equ. \eqref{equ:RASE_transform}). The solid lines are the in-phase component of the two fields, while the dashed lines are the out-of-phase components, offset vertically.}
            \label{fig:timetrace2}
        \end{figure}
 
        All results were recorded with continuous-variable-type (CV) detection. Specifically, a balanced heterodyne detection system was used to detect all output fields. The ASE/input and RASE/echo signals were recorded at the same heterodyne beat frequency, 13 MHz, by phase-coherently switching the frequency of the local oscillator during the rephasing pulses, $\pi_1$ and $\pi_2$. The $\pi_i$, $\pi_1$, and $\pi_2$ pulses were all applied to transitions at least 800 MHz detuned from the local oscillator (see Appendix B for the specific splittings), well beyond the bandwidth of the detection system. In addition to the RASE sequence pulses, three phase-reference pulses were added to the sequence, offset from the local oscillator by: 8, $-12$, and 15 MHz, to correct for trigger time jitter on the oscilloscope and interferometer phase noise on the heterodyne detection \cite{hedges10,Stuart21}. 
        
        Figure \ref{fig:timetrace2} (A) shows the heterodyne time trace of a high-gain I4LE shot. 60 \us~windows, used when analyzing each shot, are highlighted for the ASE and RASE fields along with a vacuum window (with only the local oscillator on), used as a noise reference. ASE was emitted at all times between $\pi_i$ and $\pi_1$, as seen by the increase in noise relative to the vacuum window. In this high gain regime, the correlations between the ASE and RASE fields can easily be observed by eye in a single shot, Figure \ref{fig:timetrace2} (B). There, the ASE/RASE windows have been digitally frequency-shifted to homodyne and overlapped. The RASE signal $R(t)$ was transformed to match the ASE window using
        \begin{equation} \label{equ:RASE_transform}
            R(t)' = R^\dagger(-t)/\eta \cdot e^{t_a/T} \cdot e^{i \theta}.
        \end{equation}
        This transformation takes the time-reversed complex conjugate of the RASE field, scales it by the rephasing efficiency ($\eta$), accounts for the dephasing that occurred while the field was stored on the optical transition, characterized by the write-time $T$, and adds a phase shift ($\theta$) between the ASE and RASE fields. These three parameters $\eta$, $T$, and $\theta$ were independently measured; the efficiency and phase shift were determined from the I4LE and the decay time was measured from the data in Figures \ref{fig:write_time} and \ref{fig:storage_time}. The solid lines in Figure \ref{fig:timetrace2} (B) are the in-phase component of the ASE and RASE field amplitude, while the dashed lines are the out-of-phase components. The I4LE input and echo also appear at 165 \us. The strong correlation between the ASE and RASE fields is evident in the clear overlap of the two signals.
        
\section{Results} \label{sec:results}
    Here, we present results from a series of experiments arranged into three categories: I4LE, RASE characterization, and non-classical RASE results.  

    \subsection{Inverted-four-level echo} \label{sec:results:i4le}
        To characterize the recall efficiency and storage time of the protocol and level scheme outlined in Figure \ref{fig:level_diagram}, we performed a series of I4LE experiments investigating the gain on the ASE transition, the rephasing efficiency, and the effect of the delay times ($t_a$ and $t_b$). Since the protocols are essentially identical, this approach allows us to characterize RASE performance using much brighter fields than seen in RASE itself.
        
        \begin{figure}
            \centering
            \includegraphics[width=\linewidth]{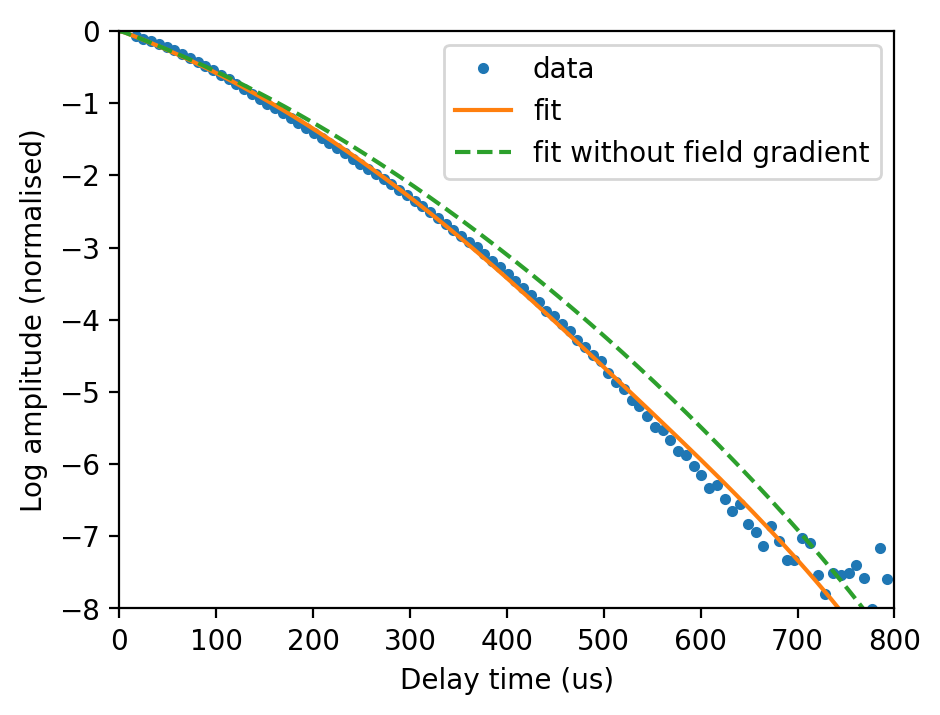}
            \caption{I4LE write-time. The echo amplitude for an I4LE is shown as a function of delay time ($2t_a + t_b)$ between the input pulse and the echo. Here $t_b$ was fixed at 0.1 \us~and $t_a$ varied from 10 to 400 \us. The fit accounts for inhomogeneous spin dephasing and a magnetic field gradient across the sample (explained in Appendix A). From the fit, we measure a 157.8 \us~write-time and predict 165 \us~without the field gradient.}
            \label{fig:write_time}
        \end{figure}
        \begin{figure}
            \centering
            \includegraphics[width=\linewidth]{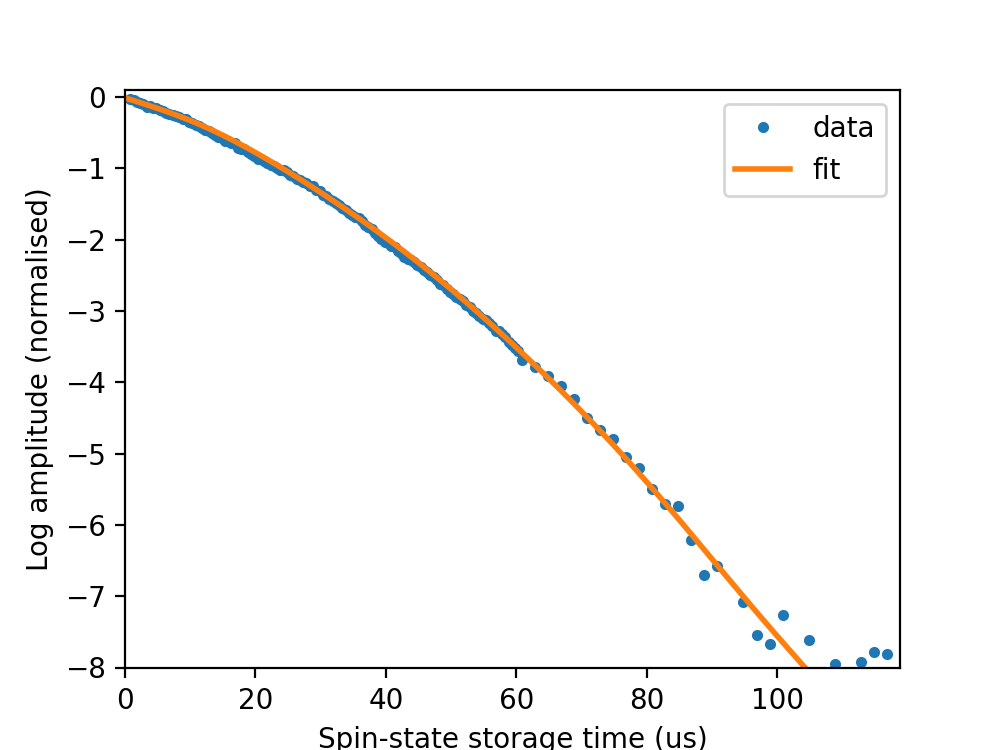}
            \caption{I4LE spin-state storage. The spin-state storage time of the I4LE was measured by changing $t_b$, the time between the two rephasing $\pi$-pulses, while $t_a$ was fixed at 10 \us. The fit accounts for inhomogeneous spin dephasing and a magnetic field gradient across the sample (explained in Appendix A). The fit gives a 25.1 \us~spin-state storage time.}
            \label{fig:storage_time}
        \end{figure}
        Figure \ref{fig:write_time} shows the echo amplitude as a function of $t_a$ for a fixed $t_b = 0.1$ \us, and Figure \ref{fig:storage_time} shows the echo amplitude as a function of $t_b$ for a fixed $t_a = 10$ \us. The $e^{-1}$ point in Figures \ref{fig:write_time} and \ref{fig:storage_time} determine the write-time and spin-state storage time of the memory, respectively. Here, write-time describes the amount of time available for encoding information on the optical transition before decoherence necessitates its transfer to the spin transition. The write-time, along with the bandwidth, determines the temporal multiplexing capacity of the memory. 

        \begin{figure}
            \centering
            \includegraphics[width=\linewidth]{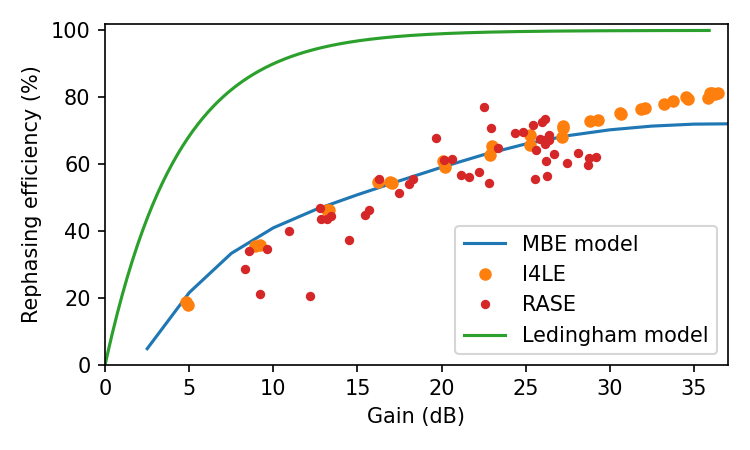}
            \caption{I4LE efficiency as a function of the gain on the input transition, $\trans[+]{5}{5}$, with delay times $t_a = 6$ \us~and $t_b = 0.1$ \us. RASE efficiencies over a smaller gain regime are also shown.
            The gain was measured by comparing the amplified input pulse to the pulse recorded without a gain medium. The Ledingham model comes from Equation \eqref{equ:RASE-eff}. The MBE model was developed to account for background absorption and imperfect rephasing.}
            \label{fig:eff}
        \end{figure}

        An upper limit on the write-time is given by the optical coherence time of 1.35 ms \cite{milosThesis} and, similarly, the storage time's upper limit is given by the hyperfine coherence time of 1.3 s \cite{Rancic16}. The decays observed are non-exponential and far faster than these limits: the $e^{-1}$ point give a storage time of 25.1~\us~and a write-time of 157.8~\us. The short decay times and non-exponential behavior in both sets of data arise from the same effect: ``inhomogeneous spin dephasing''\cite{ham03,Afzelius10,Gundogan15}, which arises when the hyperfine transition frequencies in the ground and excited state are not perfectly correlated with each other and the optical transition frequency. The decay profile seen is given by the Fourier transform of the relevant inhomogeneous line shape.  For the storage time, the decay is determined by the inhomogeneous spin dephasing on the ground state hyperfine transition. Dephasing due to this frequency distribution can be easily corrected by applying microwave rephasing pulses on this hyperfine transition, but such pulses were not applied here. For the write-time, the decay is determined by the difference in frequencies of individual ions between the ASE and RASE transitions, since the rephasing is only perfect when these are identical.  This difference is given by the difference in ground and excited state hyperfine transition frequencies. Because the dephasing arises from the difference of two transitions, rephasing pulses cannot be applied to reverse it, placing a fundamental limit on the write-time. The same limitation will arise in any other four-level quantum memory protocols, where light is recalled on a different transition to the one it was encoded on.

        In addition to inhomogeneous spin dephasing, there is some broadening due to a magnetic field gradient across the 3 mm crystal. With the crystal in the center of the superconducting magnet, the latter effect is small, contributing $\mathcal{O}(100)$~Hz of broadening (see Appendix A), and the intrinsic linewidths can be extracted by fitting the decay profiles in Figures \ref{fig:write_time} and \ref{fig:storage_time}. In both cases, the broadening is approximated by a Voigt profile. The extracted ground state linewidth of 14.8~kHz is smaller than the linewidths seen for optically-selected subgroups in other rare earth-doped \YSO\  crystals (e.g. Pr, 26 kHz \cite{Afzelius10} and Eu, 69 kHz \cite{Timoney12}), which we attribute to Er's lower nuclear magnetic sensitivity. The linewidth extracted from the write-time data is lower, 4.3~kHz, indicating a moderate degree of correlation in the excited and ground state hyperfine frequencies. 
        

        Figure \ref{fig:eff} shows the I4LE rephasing efficiency (classical memory) for a series of gains on the input/ASE transition, with $t_a = 6$ \us~and $t_b = 0.1$ \us. RASE efficiencies are also shown for the gain range from 8 to 30 dB, with 250 shots per data point. This data shows that I4LE efficiency is an accurate representation of the RASE efficiency at the same gain, with the advantage of a much higher signal requiring less acquisition time to see clear trends. In the rest of this section, we focus on the I4LE results. 
        
        The gain in Figure \ref{fig:eff}  was measured by comparing the intensity of the input pulse with, and without, $\pi_i$ applied to the sub-ensemble. The rephasing efficiency increased with gain up to a peak efficiency of 81\% at 36 dB. Also shown are two models for the efficiency data. The first (the Ledingham model) is an analytical model derived in Refs. \cite{Duda23, KateThesis} using theory presented by Ledingham \textit{et al.} \cite{Ledingham10}. This model gives the theoretical upper bound on efficiency in a two-level ensemble as a function of gain, $\alpha L$, on the ASE transition:
        \begin{equation} \label{equ:RASE-eff}
            \eta = \frac{8 \sinh^2(\frac{\alpha L}{2})}{2e^{\alpha L} -2},
        \end{equation}
        The observed efficiency will always be below this upper bound, because it assumes a homogeneous driving field (spatially and spectrally), no background absorption, no pulse errors or other loss mechanisms, no decoherence, and no dispersive effects from the gain feature.  The efficiency seen here is well below this bound for all gains. To better understand the experimental trend, we employed a more realistic, although still highly simplified, model of the system (see Appendix C). We used a numerical Maxwell Bloch Equation (MBE) model to calculate the efficiency of an inverted two-level echo on a narrow ensemble with decoherence included. We fitted two parameters, a non-unity rephasing efficiency (78\%) and non-zero background absorption (0.3 dB), to the experimental data to obtain the second fit line in Fig. \ref{fig:eff}. Here ``rephasing efficiency" is a catch-all term that includes all losses, other than background absorption, as an instantaneous decoherence event after the rephasing $\pi$-pulses. The MBE model fits the data very well in the low to medium gain regimes, relevant to quantum level measurements, and confirms that the data falls short of the theoretical upper bound due to the combination of background absorption and loss. The fitted background absorption agrees with measurements of $< 0.5$~dB. Despite having this measurement, we chose to leave the background absorption as a fitting parameter because experimental instabilities caused this absorption to vary between experiment runs.

        The model does lose accuracy at high gain, where the simplified rephasing efficiency and background absorption overcompensate for the effects of inhomogeneous driving fields. The model is spatially one dimensional, but in reality all $\pi$-pulses have a Gaussian spatial mode that is mode-matched to the collection fiber for detection. This spatial distribution means the Rabi frequency varies radially. For the inversion pulse, this results in a radially-dependent gain feature, and for the rephasing pulses, inhomogeneous rephasing. The RASE mode is defined by the effects from these pulses, and in the low gain regime, the RASE mode is well matched to the collection fiber. However, in the high gain regime, the effects of the inhomogeneous gain feature scale exponentially, leading to mode distortion and reducing the coupling efficiency to the collection fiber. Further, the refractive index of the crystal locally is dependent on gain, so a radially-dependent gain feature creates a radially dependent dispersion profile, similar to a gradient refractive index (GRIN) lens. This changes the shape of the RASE mode, which could further affect the fiber coupling efficiency. 
        
        Regardless of the additional effects operating in the high gain regime, we can make some initial inferences on the performance of the rephasing $\pi$-pulses based on Figure \ref{fig:eff}. 8\% of the observed rephasing inefficiency is explained by write-time decoherence. Thus, we bound the inversion efficiency of each rephasing $\pi$-pulse to $\geq 93\%$, if the $\pi$-pulses were the dominant source of rephasing inefficiency. 
        
        The 81\% maximum observed efficiency here includes spin-state storage and exceeds the best demonstrations observed in other solid-state systems (up to 75\%, with only ref. \cite{Schraft16} demonstrating spin-wave storage) \cite{hedges10,Jobez14,Sabooni13,Schraft16,Minnegaliev22,duranti_efficient_2024}, and approaches those seen in atomic gases (up to 95\%) \cite{Hosseini11, Hsiao18, cao20, wang19,guo25}. Although, during the submission of this paper a new Eu:\YSO paper has demonstrated 80\% \cite{mengEfficientIntegratedQuantum2026}. The substantial improvement over previous RASE demonstrations (14\% \cite{Duda23}) can be attributed to the level scheme used. All $\pi$-pulses were applied to transitions with weak oscillator strengths: 3.2\%, 7.3\%, 7.0\% ($\pi_i$, $\pi_1$, $\pi_2$) relative to the ASE transition (see Appendix B), which minimizes pulse distortion when propagating through the crystal. In the discussion, we will explain that a non-classical memory can achieve these efficiencies without operating at high gain.
        
        Further improvements to the efficiency could be achieved by two main measures. First, eliminating the remaining $\pi$-pulse inefficiencies by using composite pulses \cite{levitt86} more tolerant to pulse errors than the single, temporally square, $\pi$-pulses used here. Second, creating a gain profile with a more uniform radial distribution, either by composite pulses or complex hyperbolic secant pulses \cite{Minar10, de_seze05, Minnegaliev23} for the inversion. Reducing the background absorption during the memory preparation step with better level-cleaning protocols may also improve efficiencies, but this effect is likely to be small.


        In summary, the above results show that the level scheme used here is capable of high recall efficiencies in a free-space experiment, along with write and storage times that are sufficient for the storage of many temporal modes, investigated below. 

    \subsection{Characterization of RASE} \label{sec:classical}
        The next series of experiments characterized the performance of the RASE protocol directly. We studied correlations between the ASE and RASE windows, polarization mode mixing, and the temporal mode capacity.

        \begin{figure}
            \centering
            \includegraphics[width=\linewidth]{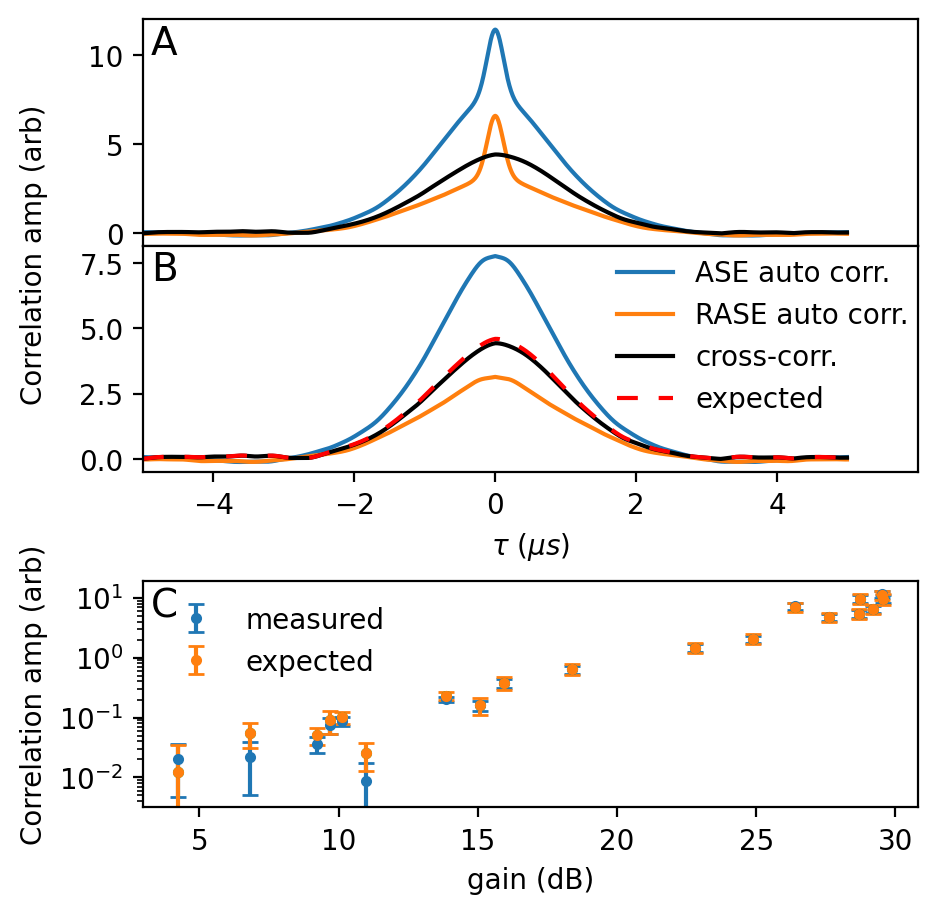}
            \caption{\textbf{A}. Shows the real component of the time-varying auto-correlation for the ASE window and the RASE windows and the magnitude of the ASE, RASE cross-correlation.. The bump in the center of the auto-correlations is the vacuum auto-correlation. \textbf{B}. shows the same graphs with the vacuum auto-correlation removed, with a time-bandwidth of 1.95 \us. The red dashed line shows the expected cross-correlation, given the amplitude of the ASE cross-correlation and the rephasing efficiency. \textbf{C}. shows the expected and measured cross-correlation amplitudes as the gain on the ASE transition is increased. \textbf{A, B} shows the gain = 25 dB data set.}
            \label{fig:corr}
        \end{figure}
        \subsubsection{Correlations}
        The time varying correlations between ASE and RASE fields were measured to investigate whether noise was present in the signals at the classical level. The correlations were studied for a series of gains between 4 and 30 dB. For each gain level, 500 RASE shots were recorded. To calculate the correlations, first the ASE and RASE windows were digitally frequency-shifted to homodyne, with a low-pass filter applied. Then, the two windows were convolved,
        \begin{align}
            C_X(\tau) &= \int A(t) \cdot R(t-\tau) dt,\label{equ:corr}\\
            C_A(\tau) &= \int A(t) \cdot A^\dagger(-(t-\tau)) dt.\label{equ:autocorr}
        \end{align}
        Equation \eqref{equ:corr} gives the time-varying cross-correlation, $C_X(\tau)$, between the ASE, $A(t)$, and the RASE, $R(t)$, time windows, while Equation \eqref{equ:autocorr} gives the auto-correlation of the ASE window with itself. A similar expression exists for the RASE auto-correlation. Time-varying correlations are complex-valued; however, the auto-correlations are always in phase and are therefore purely real-valued. The auto-correlations were used to reference the cross-correlation's amplitude back to the ASE or RASE field's amplitude. Figure \ref{fig:corr} (A) shows the auto-correlation of both fields and the cross-correlation between them. The correlations are maximal at time $\tau=0$, when the two windows are aligned. The amplitude then decays according to the bandwidth of the ensemble, with a FWHM of 1.95~\us. The narrow peaks in the center of the auto-correlations are from a vacuum auto-correlation. This has effectively, an infinite bandwidth and therefore infinitesimal temporal linewidth, but the vacuum signal is then broadened by the low-pass filter applied to the data in frequency space. In Figure \ref{fig:corr} B the vacuum auto-correlation has been removed by fitting a Gaussian and a Voigt profile to the vacuum and ASE/RASE auto correlations, respectively, and then subtracting the Gaussian from the data. Also shown in this figure is the expected cross-correlation ($C_X$) calculated from the vacuum-subtracted auto-correlations. Given that the RASE field's amplitude is the time-reversed complex conjugate of the ASE field scaled by the square root of the rephasing efficiency and a phase term, we can write the expected cross-correlation in terms of the ASE field, assuming perfectly correlated fields,
        \begin{align}
            C_X(\tau) &= \int A(t) \cdot \sqrt{\eta}A^\dagger(-(t-\tau)) e^{i\theta} dt,\\
            C_X(\tau) &= \sqrt{\eta} C_A(\tau) e^{i\theta}. \label{equ:expected}
        \end{align}
        Where $\theta$ is the phase between the ASE and RASE fields. The magnitude of the cross-correlation is then the magnitude of the ASE auto-correlation scaled by the square root of the efficiency, if the two fields are perfectly correlated. Classical noise present in the data would reduce the cross-correlation, so comparison of the expected and actual cross correlation can be used to identify classical noise. Figure \ref{fig:corr} C. shows good agreement between the expected and measured cross-correlations, across all gains, indicating a strong correlation between the ASE and RASE fields and no measurable added classical noise, within error.

        \subsubsection{Polarization mode mixing}
        \begin{figure}
        \centering
            \includegraphics[width=\linewidth]{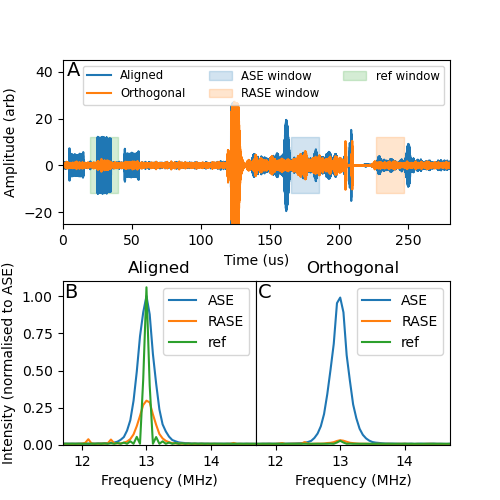}
            \caption{\textbf{A.} High gain 4-level RASE time traces, with the local oscillator's polarization aligned with (blue) and orthogonal to (orange) the light coming out of the cryostat. The ASE and RASE windows, along with one of the reference pulses, are highlighted.  \textbf{B.} and \textbf{C.}: the power spectrum of the highlighted windows for the aligned and orthogonal polarization, respectively.}
            \label{fig:polarization}
        \end{figure}    
        \begin{figure}
        \centering
            \includegraphics[width=\linewidth]{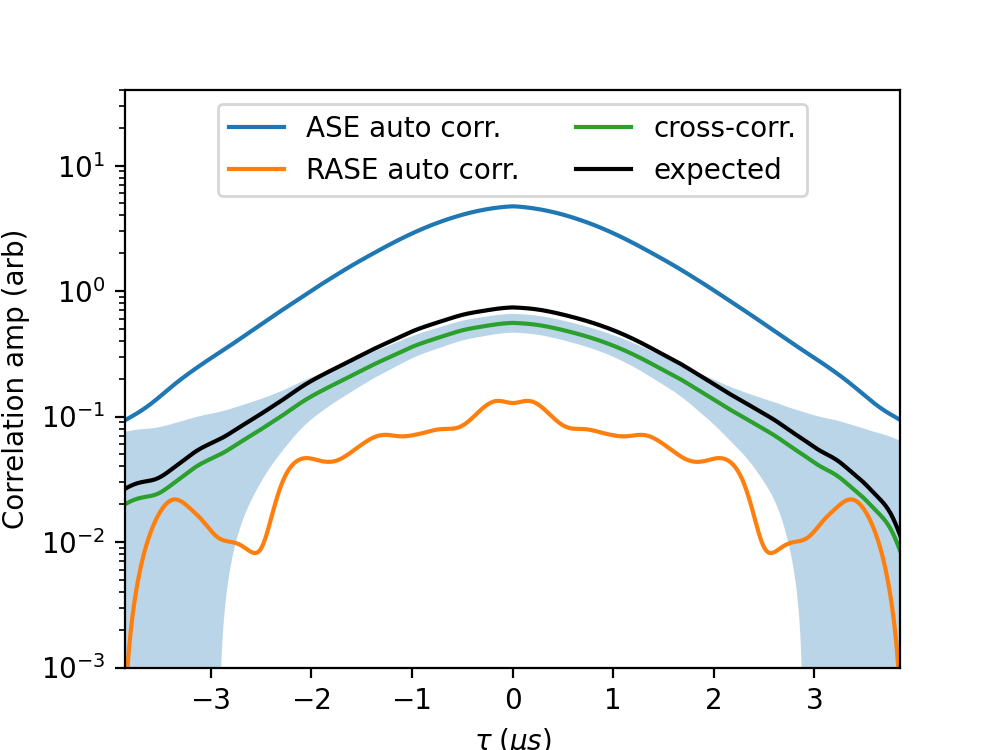}
            \caption{Classical correlations in the weak orthogonal RASE data set, shown on a log scale. The green line shows the strong orthogonal ASE - orthogonal RASE correlation, in line with the expected value, indicates that polarisation mode mixing is not making a significant contribution to the orthogonal RASE signal. The large uncertainties are due to the weak signal from this largely suppressed RASE mode.}
            \label{fig:pol_corr}            
        \end{figure}
        Polarization mode mixing is important to characterize because, if present, it would remove polarization multiplexing as a viable method for increasing memory mode capacity, and it could introduce uncorrelated ASE/RASE fields into the detection mode. \YSO, like many rare earth quantum information materials, is birefringent, so the polarization of the input light field is transformed as it propagates through the crystal in the general case. The behavior of polarized light in quantum memory protocols is not well understood, but it is possible that this effect could lead to cross-talk between polarization modes in some circumstances. 
        
        To study polarization mode mixing we recorded RASE, at maximum gain, with the heterodyne local oscillator polarization either aligned with, or orthogonal to, the polarization of the $\pi$-pulses, which define the polarization mode of the RASE signal. When the local oscillator is aligned with the $\pi$-pulses, we measure the ``aligned" signal while for a local oscillator orthogonal to the $\pi$-pulses we measure the ``orthogonal" signal. Figure \ref{fig:polarization} A. shows a time trace from both data sets, while Figures \ref{fig:polarization} B and C show the power spectrum of the ASE window, RASE window, and a reference pulse for the two polarization measurements. 
        
        The data shows that the reference pulses (which do not interact with the atoms) are emitted ($(99.1 \pm 0.4)\%$) in the same polarisation as the $\pi$ pulse, as expected. The ASE field is emitted fairly evenly in both aligned and orthogonal polarisations. This behaviour is expected; the ASE polarisation is only determined by the nature of the transition dipole moment, and our previous  echo and absorption studies have shown fairly flat polarization dependence in this propagation direction \cite{Ahlefeldt23}. In contrast, the data shows that the RASE field is predominately emitted in the polarization of the $\pi$-pulses. This can be anticipated because of the phase matching conditions for RASE recall, which are as follows. The $\pi$-pulses imprint a longitudinal phase relationship on the crystal that is dependent on the crystal birefringence, with a  wavelength given by the polarization beat length, $b_L = \lambda/\Delta n$. These $\pi-$pulses are effective at rephasing ASE signals with the same longitudinal phase relationship (i.e. the aligned ASE polarization), and somewhat ineffective at rephasing the orthogonal ASE polarization. If the $\pi$-pulses are exactly polarised along one of the principal refractive index directions (the D$_1$, D$_2$, and C$_2$ directions), there is zero rephasing of the orthogonal signal. This phase matching condition applies when the crystal is significantly longer than the polarization beat length of $\approx$\SI{100}{\micro\meter}.  

        While this partial rephasing of the \textit{orthogonal} ASE signal into is a perfectly reasonable explanation for the presence of the $(6 \pm 3)\%$ orthogonal RASE signal, we need to eliminate the possibility that part of the \textit{aligned} ASE is being rephased into the orthogonal RASE signal. If the rephasing was splitting the aligned ASE into both polarization modes in this way it would indicate polarization mode mixing was present. We can differentiate between these two sources by calculating the correlations between the fields. It was not feasible in the experiment to directly measure the correlation between the aligned ASE and orthogonal RASE, since this would require fast polarization switching. However, we can place an upper bound on the aligned ASE - orthogonal RASE correlation by considering the correlation of the orthogonal ASE - orthogonal RASE fields in Figure \ref{fig:pol_corr}. 
        The very strong correlation seen certainly indicates the orthogonal ASE field was the main source of the orthogonal RASE field, not mode mixing. Comparing the measured correlation to the expected correlation (black line) suggests that ($80\pm10$)\% of the ($6\pm3$)\% orthogonal RASE mode was correlated with the orthogonal ASE mode, limiting the polarization mode mixing to $\leq 1.2 \pm 0.6\%$. Thus, there is minimal polarization mode mixing in the experiment. The existence of the orthogonal RASE mode will not affect the RASE measurements because the heterodyne detection filters the orthogonal polarization. 
        
        \subsubsection{Temporal multiplexing capacity}
        \begin{figure}
            \centering
            \includegraphics[width=\linewidth]{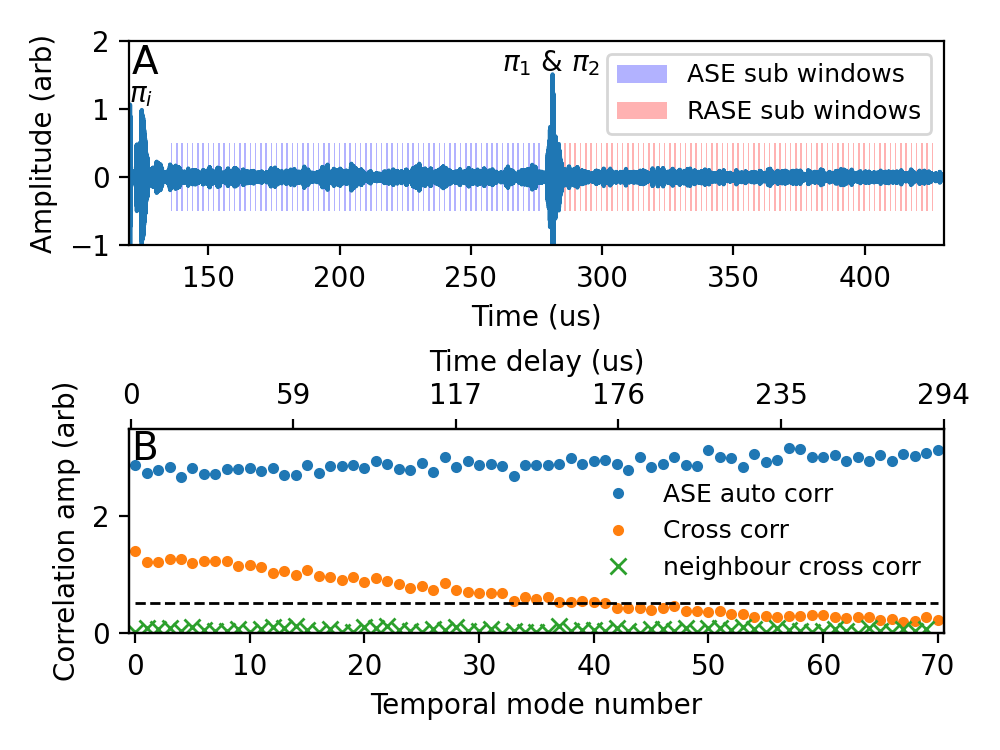}
            \caption{Data storage capacity \textbf{A}. Demonstration of 70 RASE temporal modes, achieved by splitting the ASE and RASE windows into 0.5 \us~long sub-windows separated by 2 \us. \textbf{B}. the RASE cross-correlation, ASE auto-correlation and the cross-correlation between an ASE window with the next RASE window, with the top x-axis showing the total delay time. The black dashed line indicates the cross-correlation amplitude 1/e point, intercepting with the 40th mode.}
            \label{fig:multiplex}
        \end{figure}   
        We demonstrate temporal multiplexing by subdividing the ASE and RASE temporal windows into smaller sub-windows. Figure \ref{fig:multiplex} A shows a RASE data set with a 160 \us~time gap between the inversion and rephasing $\pi$-pulses in which 70 sub-windows, each 0.5 \us~long and separated by 2 \us, are defined. Figure \ref{fig:multiplex} B shows three correlation amplitudes for each time window: the ASE auto-correlation, the cross-correlation, and the neighbor cross-correlation (the cross-correlation between the $n^{th}$ ASE window and the $(n+1)^{th}$ RASE window). The cross-correlations decay over the 158 \us~write-time, within which 40 modes are stored, 
        with the dashed line indicating the 1/e point. This gives the memory a time-bandwidth-product of 40, in agreement with the value of 39.5 that can be calculated from the write-time and memory bandwidth (from the time-varying cross-correlation, Figure \ref{fig:corr}). Finally, the near-zero neighbor cross-correlation suggests no signs of temporal mode mixing. 
        
        Taken together, the results of this section show strong correlations between the ASE and RASE fields, with no clear evidence of mode mixing. The mode purity also allowed temporal multiplexing using the full time-bandwidth-product of the memory.

        
    \subsection{Verification of quantum entanglement} \label{sec:nonclassical}

        The results discussed thus far have operated over a range of gains, with high gain used to achieve high efficiency, given by Equation \eqref{equ:RASE-eff}, and strong signals observable at the classical level. The high gain measurements also showed no evidence of classical-level noise, albeit with small, 500 shot, data sets. Non-classical tests are far more sensitive to noise. The inseparability criterion for continuous variable states \cite{Duan00} requires no greater than 1 vacuum unit of noise. Given this constant noise budget, it is common to perform non-classicality experiments with dim optical signals, near the single photon level. In the following results, a conservative gain of 7 dB was used for the same reason, with a recorded ASE signal equivalent to 2.75 vacuum noise units. In this regime, the efficiency was lower, 17\%, following the trend in Figure \ref{fig:eff}.

        The inseparability criterion for continuous variable states \cite{Duan00}, used in previous RASE demonstrations \cite{Ledingham12,kate16,Duda23}, uses a pair of Einstein-Podolsky-Rosen (EPR)-type operators,
    
        \begin{align} \label{equ:EPR_I}
            \hat{u} &= \sqrt{b}I_A + \sqrt{1-b}I_R\,,\\ \label{equ:EPR_Q}
            \hat{v} &= \sqrt{b}Q_A - \sqrt{1-b}Q_R\,.
        \end{align}
    
         I and Q are the in-phase and out-of-phase quadrature amplitudes for both fields, with the subscripts $A$ and $R$ referring to ASE and RASE, respectively. $b \in [0,1]$ is a free parameter that weights the ASE and RASE fields. The inseparability line ($\lambda$) is defined as,
    
        \begin{equation} \label{equ:insep}
            \lambda(b) = \langle (\Delta \hat{u})^2 \rangle + \langle (\Delta \hat{v})^2 \rangle.
        \end{equation}

        The inseparability criterion states that if $\lambda(b) \geq 2$ for all $b$, then the two operators are separable, and if $0 \leq \lambda(b) < 2$ for any $b$, then they are inseparable, and thus, entangled, forming a two-mode squeezed state.
        
        \begin{figure}
            \centering
            \includegraphics[width=\linewidth]{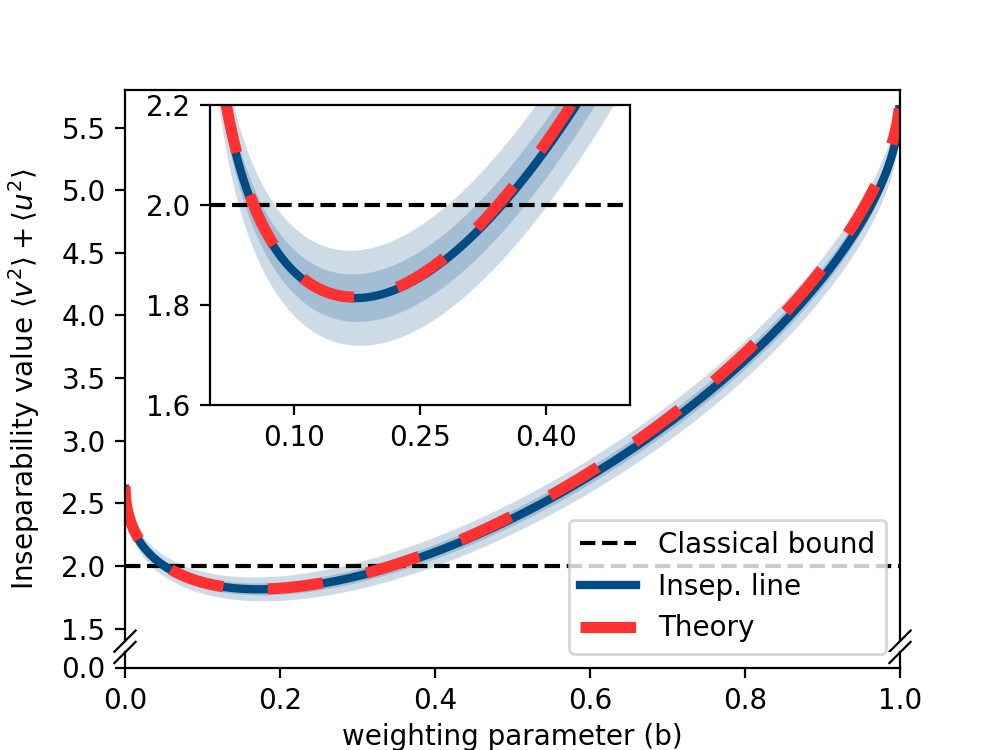}
            \caption{Inseparability criterion violation, showing a minimum of $1.81 \pm 0.05$ with the 1$\sigma$ and 2$\sigma$ confidence intervals shaded. The inset magnifies the inseparability minimum. The inseparability line dips below the classical boundary with a certainty of 3.7$\sigma$. Also shown is a theoretical model of the inseparability line, red dashed line.}
            \label{fig:insep}
        \end{figure}
    
        Figure \ref{fig:insep} shows the inseparability line, calculated from a 5,000-shot data set. Similar to the correlation measurements, the ASE and RASE windows were first digitally beaten to DC and low-pass filtered. A window length of 13.2 \us~and cut-off frequency of 280 kHz gave the best inseparability criterion violation. The inset in Figure \ref{fig:insep} zooms in on the inseparability criterion minimum, $1.81 \pm 0.05$ at $b=0.17$. The theory line, red dash-dotted, uses a model developed by Ferguson \textit{et al.} \cite{kate16}, where losses and rephasing inefficiencies are treated as beamsplitter operations that mix vacuum into the ASE and RASE fields. The model derives expressions for $\hat{u}$ and $\hat{v}$,
        \begin{align} \label{equ:insep_model}
            \langle (\Delta \hat{u})^2 \rangle &= b \big(\ell\langle I_A^2 \rangle + (1-\ell)\big) \notag \\
            & + (1-b)\Big(\ell \big(\eta\langle I_R^2 \rangle + (1 - \eta)\big) + (1 - \ell)\Big) \notag \\
            & + 2\sqrt{b(1-b)} \ell \sqrt{\eta} \langle I_A I_R \rangle,
        \end{align}
        with a similar expression for $\langle (\Delta \hat{v})^2\rangle$. Here, $\eta$ is the rephasing efficiency (17\%) and $\ell$ accounts for all losses between the crystal and the detectors (24\% through cryostat windows, 50\% from the heterodyne detection beamsplitter, and 20\% loss due to detector quantum efficiency). The inseparability criterion is violated with a certainty of 3.7$\sigma$ and shows excellent agreement with the theoretical model, indicating the absence of mode mixing other than with the vacuum state associated with losses in the experiment.

        Finally, we can infer the squeezing factor of the two-mode squeezed state was 0.4 dB at the detectors, given the inseparability minimum, 1.81. This value includes all the losses in the experiment from crystal to detector, including the heterodyne detection. We can also calculate the squeezing factor at the output of the crystal, without experimental losses, using the inseparability model to give at 1.5 dB, which is limited by the rephasing efficiency.  

\section{Discussion} \label{sec:discussion}
In this paper we made the first demonstration of quantum state storage in Er using an operating regime (temperature, magnetic field, choice of levels, and spectral preparation strategy) that can allow high performance across all quantum memory metrics relevant for both quantum repeater operation and other LOQC-based components. We showed non-classical correlations between time separated optical fields with a certainty of $3.7 \sigma$; storage efficiency up to 81\%; a storage time of 25~\us; and memory write-time of 158~\us, giving storage of 70 distinct temporal modes with a time-bandwidth product of 40.  Here, we first briefly discuss the broader applicability of the approaches and results here to other quantum memories, before discussing the prospects for further improvements in performance in this system.

The rationale behind the choice of material and operating regime here is applicable to other quantum memory approaches. As is now well recognised \cite{Rancic16}, by freezing the electron spin of a Kramers dopant, we recover (optical and nuclear spin) coherence times in Kramers ions as good as those of non-Kramers ions, with the added advantage that the electron spin freezes out the nearby nuclear spin bath (frozen core effect \cite{Zhong2014}), slowing spectral diffusion. Next, by choosing a material with large hyperfine structure relative to its inhomogeneous broadening, we can spectrally prepare the ensemble in a single hyperfine state, and then pump the selected memory ions to a spectrally well-isolated level. Thus, the memory ions are prepared in a spectral region with very low background absorption, and their spin-spin relaxation is strongly suppressed by the spin projection mismatch with the rest of the Er ensemble. Further, the bandwidth of the quantum memory in this regime is limited by the optical inhomogeneous linewidth only. Finally, by choosing a set of levels where any $\pi$-pulses see weak oscillator strengths while signal fields see strong oscillator strengths (Appendix B), we can achieve much higher memory efficiencies, even with simple temporally-square rephasing $\pi$-pulses.  This reasoning applies to \emph{any} multi-level memory protocol that uses $\pi$-pulses, and this pattern of oscillator strengths is in no way unique to $^{167}$Er:\YSO, so the approach here could profitably be used in other systems to reduce errors associated with $\pi$-pulse distortion.

We also highlighted here two important considerations for characterizing the performance of four-level quantum memory protocols. First, we described the concept of the write-time, the time available on the optical transition for encoding information. In a two- or three-level, photon-echo memory (i.e. ROSE, two- and three-level RASE), the write-time reaches the fundamental limit imposed by the coherence time. However, in this four-level system where the signal is recalled on a different optical transition to the input, a much shorter time applies, determined by uncorrelated spin inhomogeneous broadening \cite{beavan11}. The write-time is a fundamental material limit, providing further motivation to use materials with low optical inhomogeneous broadening. This write-time limit applies to all four level quantum memories, regardless of protocol. However,  note that two- and three- level non-photon-echo protocols will not reach the fundamental coherence time write-time limit seen in photon echo memories. Non-photon-echo memories, including the AFC \cite{Afzelius09} and the gradient echo memory \cite{hetet08}, have additional limitations on their write time arising from, for instance, the sharpness of the prepared spectral features used to recall the stored state.

The second consideration introduced here is the concept of polarization mode mixing, the recall of part of an input state into a polarization orthogonal to the input. Such mode mixing would limit how a memory can be used to store polarization qubits.  In a birefringent crystal like \YSO, for input light of arbitrary input polarization, polarization mode mixing is, in fact, expected, although the amount of mixing seen here was negligible for the polarization chosen. Polarization mode mixing can always be eliminated by polarizing light only along the principal axes of the refractive index tensor, indicating that some care should be taken with the polarization of light in any quantum memory in a birefringent crystal. We will discuss mode mixing, and the interaction of polarized light with birefringent quantum memories in general, in much more detail in an upcoming paper.

The RASE performance seen here is, in itself, promising for applications of $^{167}$Er quantum memories. More importantly, though, this study has allowed us to determine how to improve the critical memory parameters of fidelity, efficiency, storage time, and storage capacity, as none of these parameters are yet limited by the material properties of \isotope{Er}. Here, we use these results to lay out the pathway for further improvement of Er-based quantum memories. We will first discuss improvements to the RASE scheme specifically, and then explain how these results are relevant to a more general-purpose, read-write quantum memory. Based on the arguments above, aspects of this approach are likely to be relevant to other materials and protocols.

\textit{Fidelity} -- The inseparability condition used here to quantify the entanglement of ASE and RASE fields is not immediately translatable into fidelity, but the fact that the theoretical model, which includes no sources of infidelity, matched the data extremely well indicates the fidelity of the quantum storage step is high. The non-classical measurements support this conclusion. Infidelity arises from mixing of occupied modes into the signal, and we saw no sign of temporal mode mixing, while the polarization mode mixing was bounded above by $1.2\%$. At this stage, no improvements to the protocol are needed to improve fidelity beyond ensuring a polarization with negligible mode mixing.

\textit{Efficiency} -- For applications, the first bar for quantum memory efficiency is 90\%, because exceeding this threshold would relax the resources needed for quantum error correction \cite{muralidharan16}. In Sec. \ref{sec:results:i4le}, we showed that the remaining inefficiency in this system is due primarily to rephasing inefficiencies, with a small contribution from background absorption. There was no optimization of the pulse shape here for rephasing efficiency, but the suggested approaches in \ref{sec:results:i4le} (composite and complex hyperbolic secant pulses) have been successful in improving efficiency of $\pi$-pulses in nuclear magnetic resonance experiments and quantum memories \cite{Minar10, de_seze05, Minnegaliev23, levitt86}. Given that Ledingham's model predicts $>99\%$ efficiency for perfect pulses at high gain, $>90\%$ efficiency is likely when these measures are applied here. 

In the current setup, high efficiencies can only be reached at high gain, but placing the crystal in a Q-switchable, impedance-matched cavity \cite{Jobez14, Sabooni13} would allow the same high efficiencies in the low-gain regime \cite{williamson14} needed for quantum-level measurements. Such a setup offers further improvements to take the efficiency well beyond 90\%: the spatial mode distortion effects seen in the high gain regime can be avoided, and thinner crystals can be used, which then reduces the background absorption and any residual absorption of the rephasing $\pi$-pulses. 

Here, for our quantum-level measurements, we chose a gain of 7~dB (and resulting efficiency of 17\%) as sufficiently low to avoid classical noise. The improvements above will allow much higher efficiencies at this gain, but we also note that this was an overly conservative choice of gain, as the results of \ref{sec:classical} show minimal classical noise, even at high gain. However, a cavity-enhanced implementation would use optical depths at, or below, 7 dB, as has been used in previous cavity-enhanced memories \cite{duranti_efficient_2024, Jobez14, Sabooni13}. Operating at higher single pass gain, or in an impedance matched cavity, will push up both the quantum storage efficiency and squeezing factor achievable in an impedance-matched system.

\textit{Storage time} -- The 25~\us\ storage time here, limited by dephasing on the hyperfine transition, is sufficient for networking demonstrations over a $<5$ km network, and is equivalent to the longest delay lines used in LOQC experiments. It is straightforward to extend this storage time significantly by adding rephasing pulses on the storage hyperfine transition to the RASE pulse sequence \cite{Ma21-1hour, Heinze13}. This change allows the storage time to be extended to the $>1$~s \cite{Rancic16} hyperfine coherence time. Work to reach a storage time of 3.7~s using this approach will be reported separately. 

\textit{Mode capacity} --  The temporal mode capacity shown here, determined by the write-time, is ultimately limited by fundamental material properties: the uncorrelated inhomogeneous broadening on the different transitions employed. Improving the mode capacity, therefore, requires using other degrees of freedom. While it is experimentally complex to access, significant capacity is available in the spatial domain, given a spot size of 100~\um$^2$ compared to an average crystal size of 1~cm$^2$. Spectral multiplexing is more straightforward. Here, it would involve simply preparing multiple spectrally resolved antiholes, and running the RASE sequence on each antihole concurrently. The preparation of multiple antiholes has already been shown in this material \cite{Stuart21}. For the level scheme shown here, the optical splittings between $\pi$-pulses and between the ASE and RASE fields limits spectral multiplexing, setting an upper bound of 95 MHz (spectra shown in Appendix B). Given the 95 MHz limit, and the 158 \us~write-time, the storage of $\sim 7000$ modes is theoretically possible within a single spatial mode.


\textit{Applications of the current system} -- Since RASE combines a source and memory, it can form the basis of a quantum networking demonstration with minimal addition infrastructure. The current efficiencies and storage times are sufficient for initial demonstrations that are $\mathcal{O}$(1~km).  The underlying performance of the system is similar to, or better than, the performance seen in other recent demonstrations \cite{businger22, rakonjac_transmission_2023, knaut_entanglement_2024, liu_creation_2024, zhou_long-lived_2024}. Such a demonstration is also considerably less experimentally complicated in \isotope{Er}, because this previous work relied on frequency conversion to interact with telecom-band photons.

\textit{Read-write memory implementation} -- The source-memory combination created by RASE is well suited to repeaters as well the state generation and intial storage steps in other LOQC-based components. However,  it is not a read-write quantum memory for arbitrary states. Read-write capability is, instead, provided by closely related protocols to RASE, particularly the noiseless photon echo (NLPE)\cite{Ma21,damon11}, a four-level variant of ROSE, which can operate using exactly the same set of levels as RASE. 

RASE can be transformed into the NLPE by adding an input pulse (i.e. an I4LE), removing the inversion pulse ($\pi_i$), and changing the rephasing pulse sequence from $(\pi_1,\pi_2)$ to $(\pi_1, \pi_2, \pi_2, \pi_1)$, with the requirement that the wavevectors of $\pi_1, \pi_2$ must break the phase matching conditions that produce the echo (e.g.  by using counter-propagating pulses \cite{damon11}). Given the similarities between the two pulse sequences, and assuming good overlap of counter-propagating pulses, we expect NLPE to shown similar performance to what is reported here in terms of storage time, write-time, and lack of mode mixing. The extra pair of rephasing pulses may lower the efficiency from 81\% to $\geq 70$\%, given the $\geq 93\%$ bound on the rephasing $\pi$-pulse efficiency, assuming the memory is operated in backwards recall \cite{damon11}.  The steps described above to improve the RASE performance will have a similar effect on the NLPE.

\section{Conclusion}
We reported the first demonstration of a quantum memory protocol capable of high efficiency, fidelity and storage capacity in a telecom compatible material. We showed strong entanglement between an output state and the state retained in memory, violating the inseparability criterion by 3.7$\sigma$. The temporal multiplexing capacity (70 modes demonstrated) was the only parameter of the memory that is reaching the fundamental material limits, but the addition of spectral multiplexing provides a data capacity up to a (calculated) $10^4$ spectro-temporal modes per spatial mode. We achieved memory efficiencies up to 81\% (in the classical regime), the highest storage efficiency seen with any rare earth quantum memory protocol in a free-space , while also operating as a spin-wave memory. Modifications to the pulse sequences can improve this efficiency above 90\% in the quantum regime, and extend the 25~\us\ storage time out to the $\mathcal{O}$(1 s) coherence times. Further improvements are expected by employing an impedance-matched cavity. Such a memory could exceed the performance of fiber delay lines for LOQC and allow the demonstration of global-scale quantum networks.

\begin{acknowledgments}
This work was supported by the Australian Research Council Centre of Excellence for Quantum Computation and Communication Technology (Grant No. CE170100012) and the Commonwealth of Australia (represented by the Department of Defence) supports this research through a Defence Science Partnerships agreement.

\end{acknowledgments}

    \section*{Appendix A: Field gradient model}
        This section details a numerical model used to describe the effects of a parabolic magnetic field gradient on the write-time and storage time of the ensemble. The field gradient exists due to inhomogeneities in the 6 T magnetic field from the superconducting solenoid.     
        \begin{figure}[ht]
            \centering
            \includegraphics[width=\linewidth]{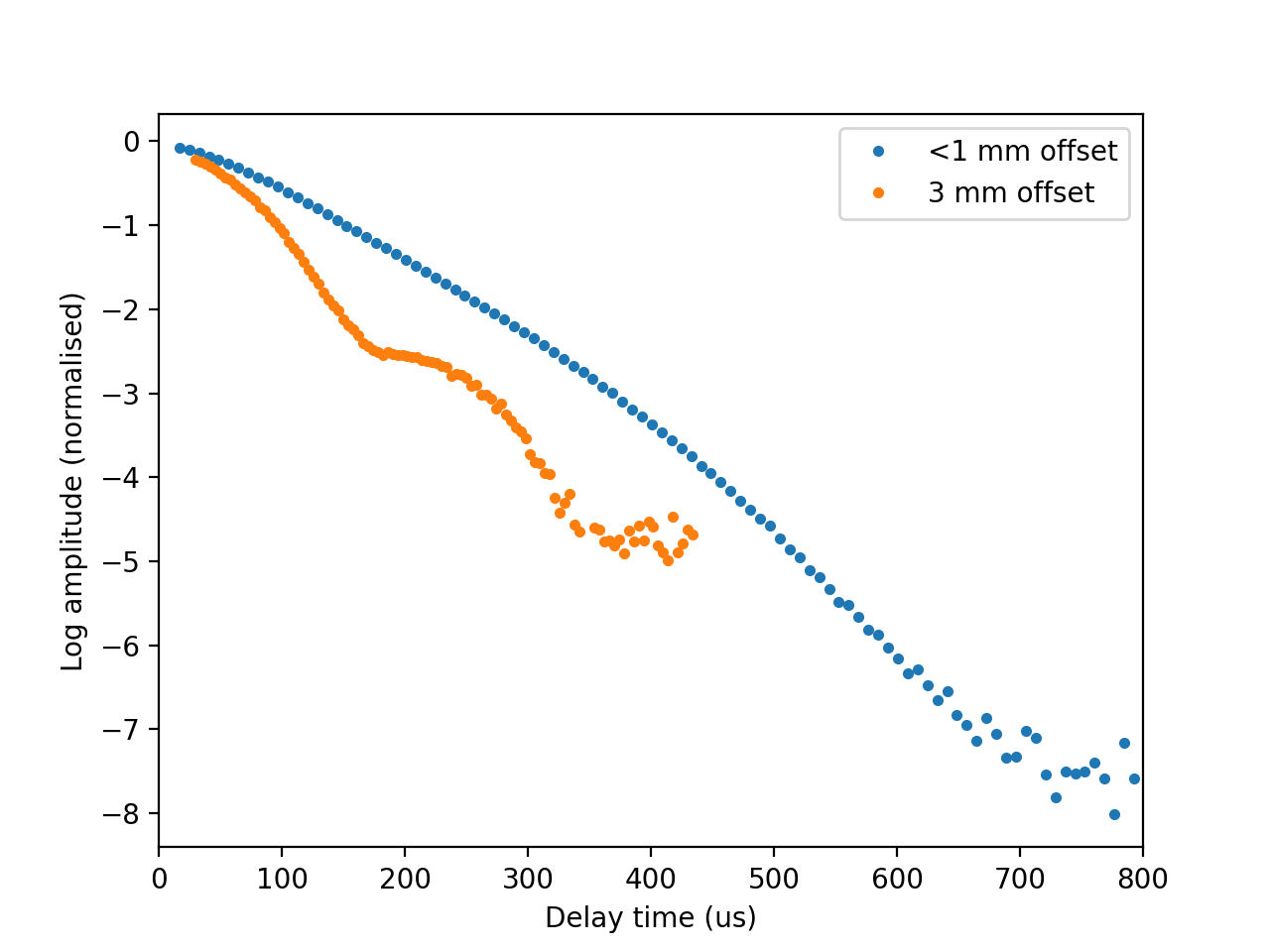}
            \caption{Two I4LE write-time data sets, with the crystal axially displaced in the magnetic field by $<1$ mm (blue) and 3 mm (orange).}
            \label{fig:field_grad:both}
        \end{figure}
        Figure \ref{fig:field_grad:both} shows two I4LE write-time measurements. The blue data set is the write-time data presented in the main text, where the crystal was within 1 mm of the magnet center. In the orange data set, the crystal was moved 3 mm, in the axial direction, from the magnet center. In this data set, the echo decay has a very pronounced modulation, arising from the magnetic field gradient across the crystal. 

        The change in magnetic field shifts the hyperfine levels of the Er ions, which causes different regions of the crystal to dephase at different rates. The dephasing leads to constructive and destructive interference between different spatial regions of the crystal.

        By measuring the magnetic field, and corresponding gradient, across the crystal we determine the Er frequency along with the density of ions at a given frequency detuning, giving an effective spectral lineshape. The echo modulation is simply the inverse Fourier transform of this lineshape. In a trivial case, a purely linear field gradient, the lineshape will be a top hat and the echo modulation follow a $|\text{sinc}|$ function. For a parabolic field gradient, as is the case here, the lineshape is numerically estimated. Finally, the decays in Figure \ref{fig:field_grad:both} are the convolution of the echo's modulation and the inhomogeneous spin dephasing \cite{ham03,Afzelius10,Gundogan15}.

        \begin{figure}[ht]
            \centering
            \includegraphics[width=\linewidth]{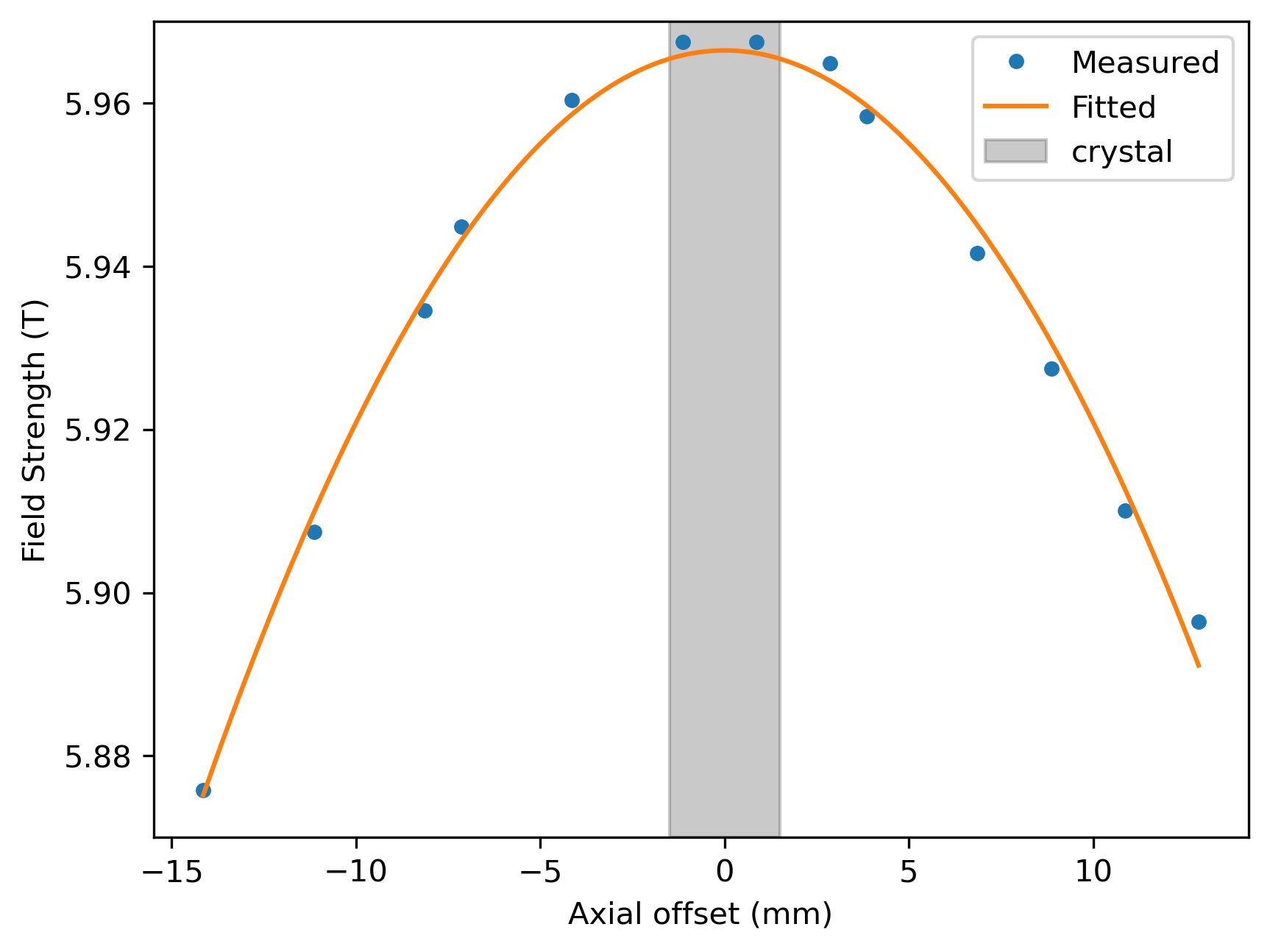}
            \caption{Measured magnetic field as a function of axial distance, taken in the center of a 6 T magnetic field. The orange line shows a parabolic fit to the data. The gray shaded region shows the position of our 3 mm thick crystal.}
            \label{fig:field_grad:field}
        \end{figure}
        Figure \ref{fig:field_grad:field} shows the magnetic field as a function of axial distance from the center of the solenoid, roughly parallel with the light propagation in the main text, with the solenoids radial direction perpendicular to the laser beam. As such, the axial and radial directions of the both the solenoid and the laser beam are equivalent. We consider field gradients along the interaction region between the crystal and the laser beam. Radial field gradients are neglected because they are significantly smaller than the axial gradients; the 30 \um~beam waist is much smaller than the 3 mm crystal length.

        The fitted parabola in Figure \ref{fig:field_grad:field} is then transformed into an atom frequency detuning by,
        \begin{align}
            B(x) &= ax^2 + bx+ c, \label{equ:field_grad:field}\\ 
            f(x) &= g \cdot B(x), \label{equ:field_grad:freq}
        \end{align}
        where $B(x)$ is the magnetic field strength, $f(x)$ is the atoms frequency detuning along the axial direction, $x$, and $g$ is the magnetic field sensitivity of the storage transition. For simplicity, we assume that the field direction does not change.
        
        We now estimate the density of ions at a given detuning frequency, $\rho(f)$, assuming the Er ions are homogeneously distributed through the crystal, $\rho(x) = c$. 

        \begin{equation}
            \rho(f) \propto \rho(x) \cdot \frac{dx}{df},
        \end{equation}        

        For a parabolic field gradient, $\frac{dx}{df}$ becomes undefined at $x=0$. To deal with this discontinuity, the crystal was divided into finite slices in the frequency domain, with the slice boundaries chosen to avoid discontinuities. 

\begin{figure}[ht]
            \centering
            \includegraphics[width=\linewidth]{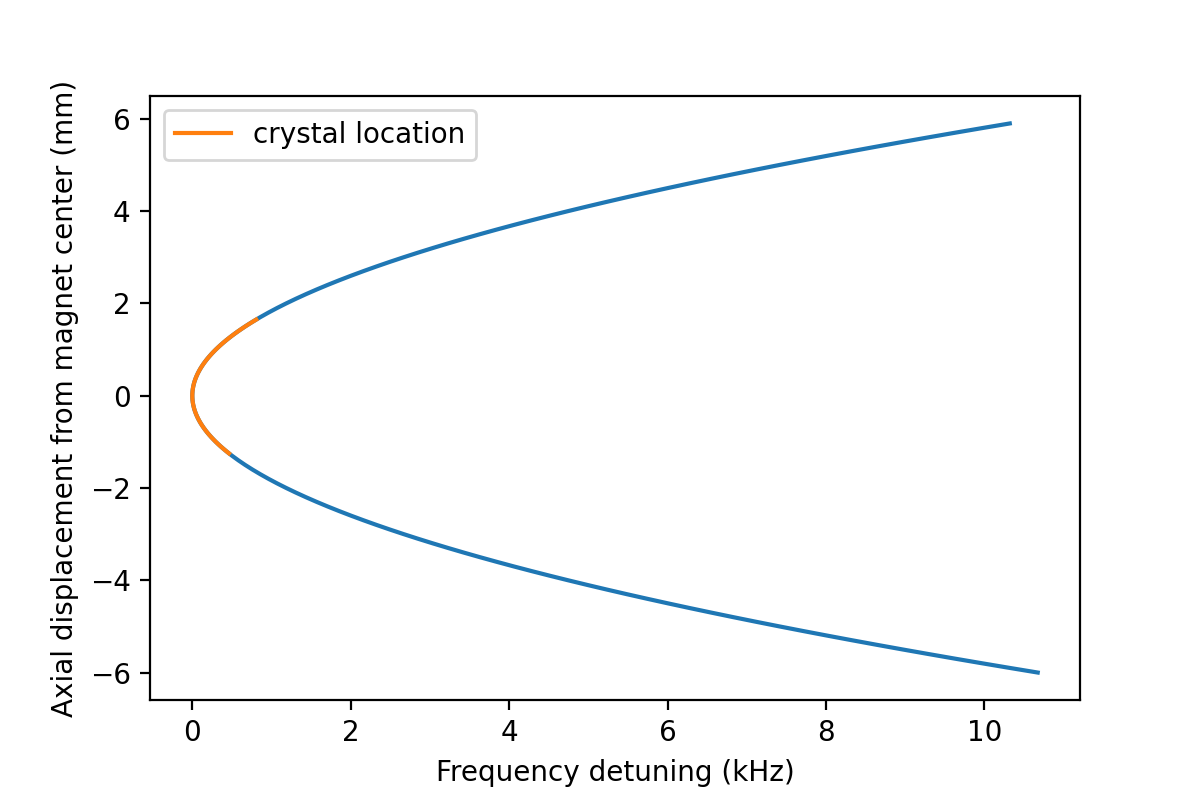}
            \caption{Frequency detuning for the write levels as the crystal is displaced axially in the magnetic field. The orange section of the line shows the crystals position in Figure \ref{fig:write_time}. Note that the parabola is slightly lopsided.}
            \label{fig:field_grad:pos_v_freq}
        \end{figure}

        Figure \ref{fig:field_grad:pos_v_freq} shows the frequency detuning as a function of axial displacement for the write-time transitions (for both data sets in Figure \ref{fig:field_grad:both}), with the crystal location for the $<1$ mm data set shown in orange.  
        %
        
        %
        \begin{figure}[ht]
            \centering
            \includegraphics[width=\linewidth]{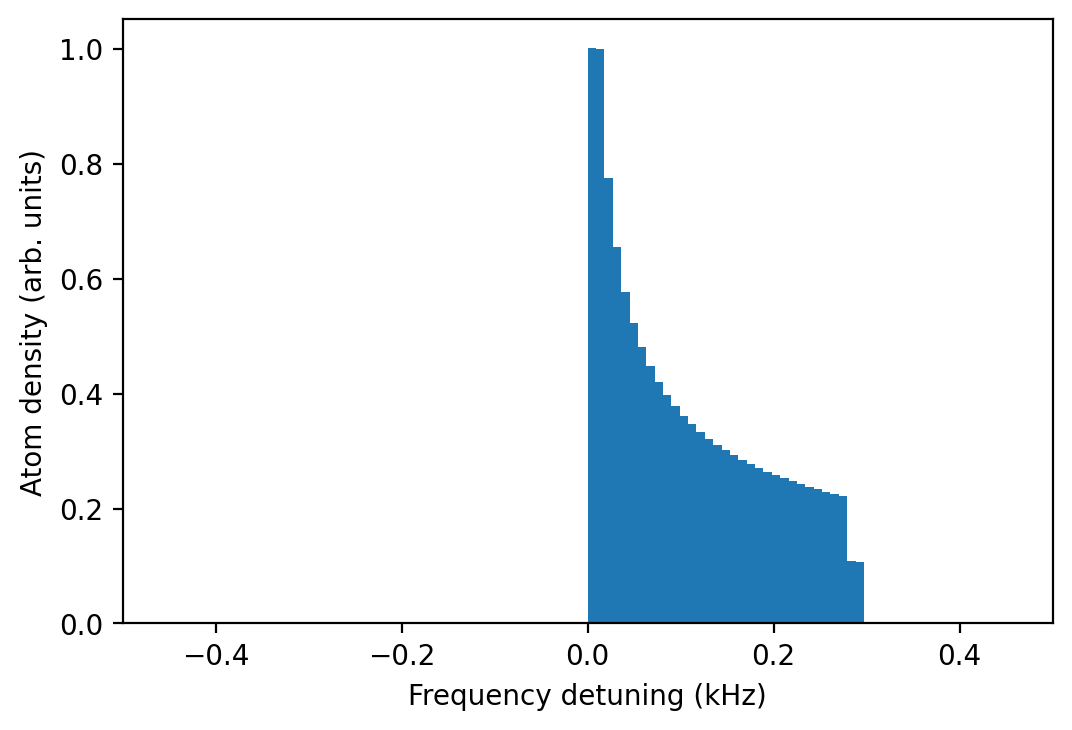}
            \caption{Ion density at different frequency detunings on the storage levels of the I4LE, due to a magnetic field gradient. For clarity, all the ions have the same optical transition frequency. There is a small step on the high frequency side of the graph, this is because the crystal was not perfectly centered in the field, so one edge side of the crystal had a slightly larger field gradient than the other side.}
            \label{fig:field_grad:density}
        \end{figure}
\begin{figure}
            \centering
            \includegraphics[width=\linewidth]{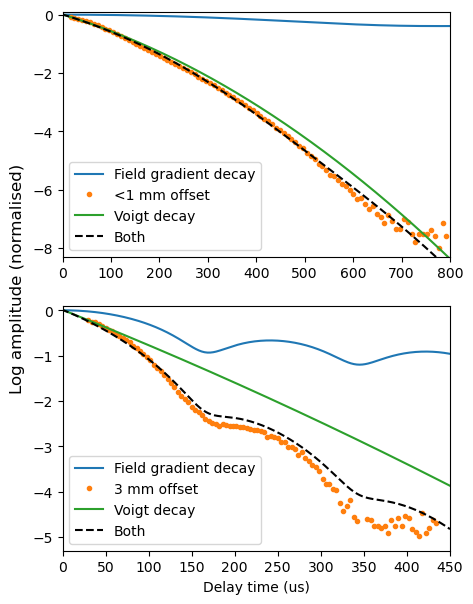}
            \caption{Fitting for both write-time data sets. Also shown is the decay due to inhomogeneous broadening (using an identical Voigt profile for both data sets) and the dephasing due to the field gradient.}
            \label{fig:field_grad:eg}
        \end{figure}

        The frequency density function, $\rho(f)$, for the $<1$ mm data is shown in Figure \ref{fig:field_grad:density}. For this data set the frequency detuning is small, at most 300 Hz, compared to the 4.3 kHz inhomogeneous spin dephasing (see main text). Thus, the effect of the field gradient is small compared to the 158 us write-time. Figure \ref{fig:field_grad:eg} shows the two write-time decay components: inhomogeneous spin dephasing and field gradient dephasing, for the $<1$ mm axial offset (top) and 3 mm axial offset (bottom), respectively.

        Ultimately, the effects of magnetic field gradients shown in the main text is rather minimal. However, this does indicate the importance of homogeneous magnetic fields, and the results shown here will be used when considering crystal size and field homogeneity. For the $< 1$ mm data, the field gradient was $\sim$ 1 mT, and shows a small, but observable effect on the write-time. 
        
        This effect can be minimized by using thin crystals, which the main text indicates that thin crystals will be used in impedance matched cavities. Further ways to minimize this effect, if needed, are: a smaller magnetic field (Ref. \cite{Rancic16} suggests that fields as low as 3 T could be used) or shimming gradient coils.

    \section*{Appendix B: Oscillator strengths and spectra}
        Here, we present thermal spectra of \isotope{Er} along with the relative oscillator strengths and relative transition frequencies of the transitions used in the RASE protocol.
        \begin{figure}[ht]
            \centering
            \includegraphics[width=\linewidth]{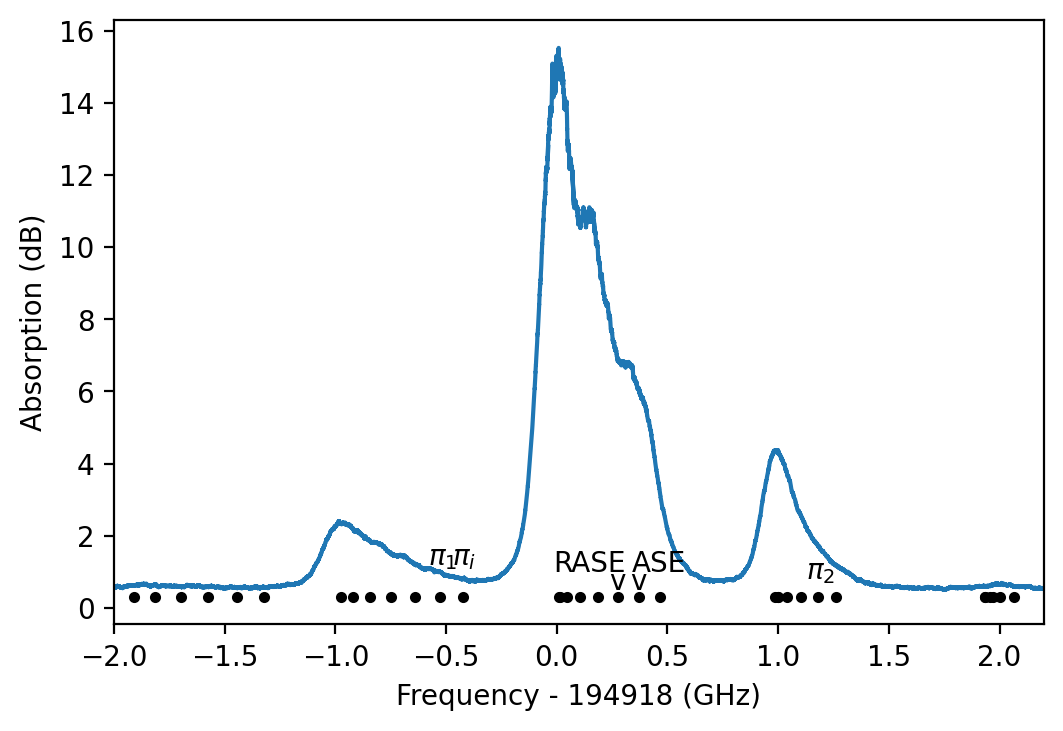}
            \caption{6 T site 2 \isotope{Er} spectra. Each black dot represents an optical transition, with the ASE, RASE, and $\pi$-pulse transitions labelled.}
            \label{fig:spectra}
        \end{figure}
        Figure \ref{fig:spectra} shows the spectra of \isotope{Er} in thermal equilibrium, with an inhomogeneous linewidth of 150 MHz \cite{Rancic16}. The black dots each indicate a unique optical transition and the transitions used in the RASE transition are labelled $\pi_i$, $\pi_1$, $\pi_2$, ASE, and RASE. This figure can be used to understand the limitation of the bandwidth of the photon memory. The ultimate limit of bandwidth of a memory is the optical inhomogeneous linewidth, here 150~MHz, but there can be lower protocol-dependent limits. Here, the bandwidth is limited by off-resonant excitation of other transitions during any step of the pulse sequence, as this will cause errors. For this particular level scheme, the ASE and RASE transitions are the closest, separated by 95 MHz, which defines the maximum memory bandwidth.
        \begin{table}[ht]
        \begin{tabular}{r|r|r}
        $\pi$-pulse/ & Relative oscillator & Freq. (MHz) \\ 
        transition   & strength (\%)       & - 194918 GHz \\ \hline
        $\pi_i$    & 3.2~                 & ~-430.36 \\              
        ASE        & 89.6~                & ~~363.75\\    
        LO (ASE) & ---             & ~~348.75\\
        $\pi_1$    & 7.3~                 & ~-538.14 \\              
        $\pi_2$    & 7.0~                 & ~~1170.28\\               
        RASE       & 81.7~                & ~~268.33\\
        LO (RASE) & ---             & ~~253.33\\
        \end{tabular}
        \caption{Relative oscillator strengths and transition frequencies of the transitions used in the 4-level RASE pulse sequence, along with the local oscillator's frequency. In the RASE protocol the ASE and RASE transitions are the closest to each other, split by 95.42 MHz.}
        \label{tab:osc_str}
        \end{table}   

        Table \ref{tab:osc_str} shows the relative oscillator strengths and transition frequencies of the transitions used in the 4-level RASE pulse sequence. As discussed in the text, the $\pi$-pulse transitions are all significantly weaker than the ASE and RASE transition, allowing the gain on the two output transitions to be high while the $\pi$-pulses experience low absorption and thus minimal distortion.

\section*{Appendix C: Maxwell Bloch model of memory efficiency}
To better model the I4LE efficiency, Fig. \ref{fig:eff}, we employed a simplified, one dimensional Maxwell Bloch model, using the standard assumptions. We considered an ensemble of two-level atoms (states $\ket{a}$ and $\ket{b}$) driven by a classical light field $E = E_0e^{i(kz-\omega_0t)}$ in the weak interaction limit such that the rotating wave approximation applied. The light polarisation was ignored and the (electric) dipole transition moment $d_{ab}$ was chosen to match the 36~dB peak absorption for this 6 mm, 0.005\% Er sample. The atoms were represented by a density matrix  and relaxation was included via phenomenological decay constants $T_1 = 10$~ms and $T_2 = 1$~ms (which had minimal effect over the short, $
\mathcal{O}$(\SI{1}{\micro\second}), delay times modeled). This resulted in optical Bloch equations in the rotating frame:
\begin{align}
\dot{\rho}_{aa} &= -\frac{i}{2}\left[\rho_{ba}\chi^*-\rho_{ab}\chi \right] +\frac{1}{T_1}(1-\rho_{aa}),\nonumber\\
\dot{\rho}_{bb} &= +\frac{i}{2}\left[\rho_{ba}\chi^*-\rho_{ab}\chi \right] -\frac{1}{T_1}\rho_{bb},\nonumber\\
\dot{\rho}_{ab} &= -\frac{i}{2}(\rho_{bb}-\rho_{aa})\chi^*+i\Delta\rho_{ab} -\frac{1}{T_2}\rho_{ab},\nonumber\\
\label{masterd}
\end{align}
where $\chi = -\frac{1}{\hbar}d_{ab}E_0e^{ikz}$ is the Rabi frequency and $\Delta$ is the atomic detuning. The electric field generated by the atoms was calculated by adding a source term,
$$P = N\Tr(d_{ab}\rho),$$
where $N$ is the atomic number density, to Maxwell's wave equation:
\begin{equation}
    \nabla^2E-\frac{n^2}{c^2}\frac{\partial^2 E}{\partial t^2} =\mu_0\frac{\partial^2 P}{\partial t^2}.\label{Mwaveeq}
\end{equation}
Here, $\mu_0$, $n$, and $c$ are the free-space permittivity, refractive index, and speed of light respectively. Applying the usual slowly varying amplitude approximation and translating to the co-moving frame $z=\xi$ and $\tau =\left(t-\frac{nz}{c}\right)$ allows expressing this in terms of a single variable,

\begin{align}
    \frac{\partial E}{\partial \xi}+\mathrm{c.c.} = \frac{i\mu_0 N\omega c}{n} d_{ab}^*\rho_{ab}+\mathrm{c.c.}.
\end{align}
The crystal was divided into optically thin slices (0.5 mm thick) and the above equations were used to calculate the electric field generated at each slice, which was then fed in as the input field for the following slice. 

To model the I4LE, we considered a 400~kHz-wide atomic ensemble with an initial inversion ranging from 0 (0 dB gain) to 1 (36 dB gain). This ensemble was driven by a weak input pulse ($<1\%$ of a $\pi$-pulse), and then, after a short wait time (7 \us), a hard $\pi$-pulse. The I4LE efficiency was calculated by comparing the resulting echo to the input pulse. Losses were modelled with a single rephasing efficiency parameter that attenuated $\rho_{ab}$ after the $\pi$-pulse. Background absorption was included as a reduction of gain in each crystal slice, after the $\pi$-pulse. These parameters were fit to the experimental data to provide the curve in Fig. \ref{fig:eff}. To efficiently fit the data, the fitting parameters were bounded to reasonable values. The background absorption and rephasing efficiency was bounded between 0 - 3 dB and the fit was weighted to prioritise the low-mid gain regime (0 - 20 dB), as the model breaks down in the high gain regime.
\bibliography{bib}  
\end{document}